\documentclass[11pt,twoside]{article}
\usepackage[ansinew]{inputenc}
\usepackage{amsthm}
\usepackage{amssymb,bm}
\usepackage{amsmath}
\usepackage[english]{babel}
\usepackage{array}
\usepackage{amsfonts}
\usepackage{latexsym}
\usepackage{longtable}
\numberwithin{figure}{section}
\numberwithin{table}{section}

\usepackage[pdftex]{color,graphicx}
\usepackage[usenames,dvipsnames,table]{xcolor}
\usepackage{float} 
\usepackage{subfigure} 
\usepackage{wrapfig} 
\usepackage[rflt]{floatflt} 
\graphicspath{{./FiguresAPoissST/}}

\usepackage{longtable} 
\usepackage{multirow}
\usepackage{multicol}
\usepackage{pdflscape} 
\usepackage{booktabs}
\usepackage{listings}
\usepackage{textcomp}
\usepackage{algorithm}
\usepackage{algorithmicx}
\usepackage{algpseudocode}
\date{}
\numberwithin{equation}{section}
\numberwithin{figure}{section}

\def \CC {\hbox{\boldmath$C$}}

\def \HH {\hbox{\boldmath$H$}}
\def \II {\hbox{\boldmath$I$}}

\def \WW {\hbox{\boldmath$W$}}

\def \ee {\hbox{\boldmath$e$}}
\def \ff {\hbox{\boldmath$f$}}

\def \uu {\hbox{\boldmath$u$}}
\def \vv {\hbox{\boldmath$v$}}
\def \xx {\hbox{\boldmath$x$}}
\def \yy {\hbox{\boldmath$y$}}

\def \cero {\hbox{\boldmath$0$}}

\def \bbeta {\hbox{\boldmath$\beta$}}

\def \ttheta {\hbox{\boldmath$\theta$}}

\def \eepsilon {\hbox{\boldmath$\epsilon$}}

\def \GGamma {\hbox{\boldmath$\Gamma$}}

\newcommand{\R}{\mathbb{R}}
\newcommand{\PR}{\mathbb{P}}

\newcommand{\abs}[1]{\lvert#1\rvert}

\title{Area-level spatio-temporal Poisson mixed models for predicting domain counts and proportions\footnote{Supported by the Instituto Galego de Estat\'{\i}stica,
by the grants MTM2017-82724-R and PGC2018-096840-B-I00 of the Spanish Ministerio de Ciencia, Educaci\'on y Universidades and by the
Xunta de Galicia (Grupos de Referencia Competitiva ED431C-2016-015 and Centro Singular de Investigaci\'on de Galicia ED431G/01), all of them through the ERDF.}}
\author{Miguel Boubeta$^{1}$,  Mar\'{\i}a-Jos\'e Lombard\'{\i}a$^{2}$, Francisco Marey-P\'{e}rez$^{3}$, Domingo Morales$^{4}$
\vspace{0.1 cm}\\
{\small$^{1}$ Universidade da Coru\~na, Spain,}\\
{\small$^{2}$ Universidade da Coru\~na, CITIC, Spain,}\\
{\small $^{3}$Universidad de Santiago de Compostela, Escuela Polit\'ecnica Superior de Ingenier\`{\i}a, Spain},\\
{\small$^{4}$ Universidad Miguel Hern\'{a}ndez de Elche, Centro de Investigaci\'on Operativa, Spain.}\\
{\small July 23, 2020}}
\begin{document}
\maketitle

\vspace*{-10mm}
\begin{abstract}
This paper introduces area-level Poisson mixed models with temporal and SAR(1) spatially correlated random effects.
Small area predictors of the proportions and counts of a dichotomic variable are derived from the new models and the corresponding mean squared errors are estimated by parametric bootstrap.
The paper illustrates the introduced methodology with two applications to real data.
The first one deals with data of forest fires in Galicia (Spain)  during 2007-2008 and the target is modeling and predicting counts of fires.
The second one treats data from the Spanish living conditions survey of Galicia of 2013 and the target is the estimation of county proportions of women under the poverty line.
\vspace{0.1 cm}
\end{abstract}

%
%
\section{Introduction}\label{sec1}
%
%
This paper introduces statistical methodology for estimating counts, in particular, problems as diverse as the counts of forest fires and poverty proportions in areas of Galicia.
Galicia is an autonomous community in the northwest of Spain with an economic activity strongly related to natural resources.
During the last years, a problem that the local government faces and that directly attacks natural resources are fires.
Year after year, the number of forest fires and the area burned in Galicia has been greater than in other regions of Spain.
The objective is to study the number of wildfires by forest areas and months, so that the government can know their behavior in each area and take appropriate measures in each.

Another field of interest to apply this methodology is poverty studies.
This work estimates the poverty map for Galician women, county by county. The increase in socioeconomic differences between the areas of the inner zone, which are poorer,
and the coastal areas, that have greater development, is worrying.

Therefore, developing statistical methodologies to model counting data and to predict domain counts and proportions is important to understand
the phenomena studied (like poverty or arsons) and, consequently, to make decisions about public policies.
This manuscript introduces several extensions of the basic area-level Poisson mixed model,
for better fitting the needs of real data and giving rise to increasingly complex and realistic models.

When auxiliary variables related to the target count variable are available at the area level, the Poisson mixed models investigated by Boubeta et al. (2015, 2016, 2017, 2019),
the multinomial logit mixed models studied by L\'opez-Vizca\'{\i}no et al. (2013, 2015) or the compositional mixed model given by Esteban et al. (2020)
link all the domains to enhance the estimation at a particular area, that is, they borrow strength from other areas.
Their models have random effects taking into account the between-domain variability that is not explained by the auxiliary variables,
but they assume that the domain random effects are independent. However, in
socioeconomic, environmental and epidemiological applications, estimates for areas that are
spatially close may be more alike than estimates for areas that are further apart. In fact,
Cressie (1993) shows that not employing spatial models may lead to inefficient inferences when
the auxiliary variables does not explain the spatial correlation of the study variable.

In small area estimation (SAE), modelling the spatial correlation between data from different areas allows to borrow even more strength from the areas.
This recommendation was applied to the basic Fay-Herriot model by Singh et al. (2005). Later, several authors have proposed
new spatial area-level linear mixed models.
Petrucci and Salvati (2006), Pratesi and Salvati (2008), Molina et al. (2009), Marhuenda et al. (2013) and Chandra et al. (2015) consider linear mixed models (LMM) that extend the
Fay-Herriot model.  Most of these papers assume that area effects follow a simultaneously autoregressive process of order 1 or SAR(1).

In the Bayesian framework, Moura and Migon (2002) and You and Zhou (2011) consider spatial stationary mixed models, Sugasawa et al. (2015) study an empirical Bayesian estimation method
with spatially non-stationary hyperparameters for area-level discrete and continuous data having a natural exponential family distribution.
Choi et al. (2011) examine several spatio-temporal mixed models in small area health data applications and develop new accuracy measures to assess the recovery of true relative risks.
They apply the spatio temporal models to study chronic obstructive pulmonary disease at county level in Georgia.

Concerning nonparametric and robust methods, Opsomer et al. (2008) give a small area estimation procedure using penalized spline regression with applications to spatially correlated data.
Ugarte et al. (2006) and Ugarte et al. (2010) study the geographical distribution of mortality risk using small area techniques and penalized splines.
Chandra et al. (2012) introduce a geographical weighted empirical best linear unbiased predictor for a small area average and give an estimator of its conditional mean squared error (MSE).
Baldermann et al. (2016) describe robust SAE methods under spatial non-stationarity linear mixed models.
Chandra et al. (2017) introduce small area predictors of counts under a non-stationary spatial model.
Chandra et al. (2018) develop a geographically weighted regression extension of the logistic-normal and the Poisson-normal generalized linear mixed models (GLMM) allowing for spatial nonstationarity.

A spatio-temporal extension of the Fay-Herriot model was proposed by Singh et al. (2005) using the Kalman filtering approach.
Under this model, they obtain a second order approximation to the MSE of the EBLUP.
Later, Pereira and Coelho (2012) and Marhuenda et al. (2013) derive empirical best linear unbiased predictors under similar models.
Specifically, they consider a SAR(1) spatial correlation structure and an AR(1) process for the temporal component.
Esteban et al. (2012, 2016) present new spatio-temporal model by assuming AR(1) and MA(1)-correlated random effects.
They also propose bootstrap procedures for estimating the mean squared error and they analyse the behaviour of the proposed model against other simpler models through several simulation experiments.

The above cited papers introduce SAE procedures that borrows strength from spatial or temporal correlations.
They mainly apply spatial or temporal temporal LMMs to the small area estimation setup.
However, few of them deals with empirical best predictors (EBP) under spatial or spatio-temporal GLMMs.
This work partially covers that gap and studies an area-level Poisson mixed model containing SAR(1) spatially correlated domain effects and independent time effects.
The final target is the estimation of domain counts and proportions.

This paper is organized as follows.
Section \ref{sec2} introduces the new area-level Poisson mixed model and several particularizations and limit cases.
This section derives an algorithm to calculate the method of moment estimators of the model parameters and
presents three bootstrap algorithms for testing the significance of the variance and autocorrelation parameters.
Section \ref{sec3} gives the empirical best predictor of the Poisson parameter and of the domain and domain-time random effects.
This section  proposes a parametric bootstrap procedure to estimate the MSEs of the EBPs.
Section \ref{sec.firedata} illustrates the developed methodology in a environmental field. The target is to predict counts of fires by forest areas of Galicia.
Section \ref{sec.povdata} estimates women poverty proportions by counties of Galicia.
Section \ref{sec.conclus} collects the main conclusions.
The paper has two appendixes.
Appendix A contains the mathematical derivations of the fitting algorithm.
Appendix B investigates the behaviour of the proposed fitting algorithm and empirically compares the performance of the plug-in and EBP by means of simulation experiments.
%
%
%
%
\section{The model}\label{sec2}
%
%
This section extends the area-level Poisson mixed model, introduced by Boubeta et al. (2016), to the spatio-temporal context.
In particular, the section introduces a model with SAR(1)-correlated domain random effects and with independent domain-time random effects.
The new model is a generalization of Model 1, investigated by Boubeta et al. (2017).

Let $D$ and $T$ be the total number of domains and time instants respectively, with the corresponding indices $d=1,\ldots, D$, and $t=1,\ldots, T$.
Consider a spatio-temporal model with two independent vectors
$$
\vv_{1}=\underset{1\leq d \leq D}{\mbox{col}}(v_{1,d})\quad\mbox{ and }\quad
\vv_{2}=\underset{1\leq d \leq D}{\mbox{col}}(\underset{1\leq t \leq T}{\mbox{col}}(v_{2,dt})),
$$
containing the domain and the domain-time random effects respectively.
The model assumes that the vector of domain random effects $\vv_1$ is spatially correlated, following a SAR(1) process
with unknown autoregression parameter $\rho$ and known proximity matrix $\WW$, i.e.
$$
\vv_1=\rho \WW\vv_1+\uu_1,
$$
where $\uu_1\sim N_D(\cero,\II_D)$, $\cero$ is the $D\times 1$ zero vector and $\II_D$ denotes the $D \times D$ identity matrix.
It also assumes that the matrix $(\II_D-\rho \WW)$ is non-singular. Then, $\vv_1$ can be expressed as
\begin{equation}\label{sec2-SAR}
\vv_1=(\II_D-\rho \WW)^{-1}\uu_1.
\end{equation}
For the proximity matrix $\WW$, we assume that it is row stochastic.
Then, the autoregression parameter $\rho$ is a correlation, $\rho \in (-1,1)$, and is called spatial autocorrelation parameter.
Some of the most used proximity matrices are based on: (i) common borders, (ii) distances and (iii) $k$-nearest neighbours. In all cases, the proximity matrix $\WW$ is obtained from an original proximity matrix $\WW^0$ with diagonal elements equal to zero and remaining entries depending on the employed option.
In option (i), the non diagonal elements of $\WW^0$ are equal to 1 when the two domains corresponding to the row and the column indices are regarded as neighbours and zero otherwise. In Option (ii), the nondiagonal elements of the proximity matrix $\WW^0$ are defined by applying a monotonously decreasing function to the domain distances; for example, by using the inverse function. Finally, the non diagonal elements of $\WW^0$ in option (iii) are 1 if they correspond to the  $k$-nearest neighbours of a given domain and zero otherwise. For each option, the row standardization is carried out by dividing each entry of $\WW^0$ by the sum of the elements in its row. Consequently, $\WW$ is row stochastic.
Equation (\ref{sec2-SAR}) implies that $\vv_1\sim N_D(\cero,\GGamma(\rho))$, where
\begin{equation}\label{sec2-sarmatrix}
\GGamma(\rho)=\big(\gamma_{d_1d_2}(\rho)\big)_{d_1,d_2=1,\ldots,D}=\CC^{-1}(\rho)
\end{equation}
and $\CC(\rho)=(\II_D-\rho \WW)^{\prime}(\II_D-\rho \WW)$.
Equation (\ref{sec2-SAR}) implies that $\vv_1=\underset{1\leq d \leq D}{\mbox{col}}(v_{1,d})\sim N_D(\cero,\GGamma(\rho))$, where $\GGamma(\rho)$ is given in (\ref{sec2-sarmatrix}). Therefore, the density function of the domain random effects $\vv_1$ is
\begin{equation*}
f_v(\vv_1)=(2\pi)^{-D/2}|\GGamma(\rho)|^{-1/2}\exp\left\{-\frac12\vv_1^\prime\GGamma^{-1}(\rho)\vv_1\right\}.
\end{equation*}
Further, it holds that
$v_{1,d}\sim N \big(0,\gamma_{dd}(\rho)\big)$ and $v_{1,d_2}|v_{1,d_1}\sim N \big(\mu_{d_2|d_1},\sigma^2_{d_2|d_1}\big)$, where
\begin{equation*}
\mu_{d_2|d_1}=\frac{\gamma_{d_1d_2}(\rho)}{\gamma_{d_1d_1}(\rho)}\,v_{d_1},\quad
\sigma^2_{d_2|d_1}=\gamma_{d_2d_2}(\rho)-\frac{\gamma_{d_1d_2}^2(\rho)}{\gamma_{d_1d_1}(\rho)}.
\end{equation*}
The interaction domain-time random effects, $\vv_2$, are assumed to be independent over time, i.e.
\begin{equation*}
\vv_{2d}=\underset{1\leq t \leq T}{\mbox{col}}(v_{2,dt})\sim N(\cero,\II_T),\quad
\vv_{2}=\underset{1\leq d \leq D}{\mbox{col}}(\vv_{2d})\sim N(\cero,\II_{DT}).
\end{equation*}
Then, the join density function of the random effects $\vv_1$ and $\vv_2$ is
$$
f_v(\vv_1,\vv_2)=(2\pi)^{-D(T+1)/2}|\GGamma(\rho)|^{-1/2}\exp\left\{-\frac12\,\vv_1^\prime\GGamma^{-1}(\rho)\vv_1-\frac12\,\vv_2^\prime\vv_2\right\}.
$$
The distribution of the target variable $y_{dt}$, conditionally on the random effects $v_{1,d}$ and $v_{2,dt}$, is
\begin{equation*}
y_{dt}\vert v_{1,d},v_{2,dt}\sim\mbox{Poisson}(\mu_{dt}),\quad d=1,\ldots,D,\,\,t=1,\ldots,T,
\end{equation*}
where $\mu_{dt}$ denotes the mean of the Poisson distribution.
We assume that $\mu_{dt}$ can be expressed as $\nu_{dt} p_{dt}$, where $\nu_{dt}$ is a known natural number.
In the application to fire data, we take $\nu_{dt}=1$ and $\mu_{dt}=p_{dt}$.
In the application to poverty data, $y_{dt}$ counts the number of sampled poor women and $\nu_{dt}$ is the corresponding sample size of women in domain $d$ and time instant $t$. In that case, $\nu_{dt}$ and $p_{dt}$ can be interpreted as size and probability parameters respectively.
The advantage of using a Poisson model instead of a binomial model is that we can avoid the calculation of combinatorial numbers with values outside the computer range. Besides, by the nature of our poverty data problem, $\nu_{dt}$ takes large values and $p_{dt}$ small values. Then, everything points to a good behavior of the Poisson model in addition to its computational advantages.

As $\nu_{dt}$ is assumed to be known, the Poisson parameter, $\mu_{dt}$, is determined if and only if one knows the parameter $p_{dt}$.
In what follows, we will refer to $p_{dt}$ as target parameter.
To define the area-level Poisson mixed model with SAR(1) spatial domain effects and independent time effects,
we express the natural parameter $\log \mu_{dt}$ in terms of a set of $p$ covariates, i.e.
\begin{align*}
\mbox{Model ST1: } \log\mu_{dt}&=\log\nu_{dt}+\log p_{dt}\\
&=\log\nu_{dt}+\xx_{dt}\bbeta+\phi_1 v_{1,d}+\phi_2 v_{2,dt},\quad d=1,\ldots,D,\,\,t=1,\ldots,T,
\end{align*}
where $\mu_{dt}= \mathbb{E}[y_{dt}|v_{1,d},v_{2,dt}]$, $\xx_{dt}=\underset{1\leq k \leq p}{\hbox{col}^\prime}(x_{dtk})$ is the row vector of auxiliary variables,
$\bbeta=\underset{1\leq k \leq p}{\hbox{col}}(\beta_k)$ is the column vector of regression coefficients and $\phi_1$ and $\phi_2$ are standard deviation parameters.
Further, Model ST1 assumes that the $y_{dt}$'s are independent conditionally on the random effects $\vv_1$ and $\vv_2$.
If we define $u_{1, d}=\phi_1 v_{1, d}$ and $u_{2, dt}=\phi_2 v_{2, dt}$, then $\phi_1^2$ and $\phi_2^2$ are variance component parameters for $u_{1, d}$ and $u_{2, dt}$ respectively.
They can be interpreted as variabilities between domains and between time periods within each domain respectively.
Particular or limit cases of Model ST1 are:
\begin{itemize}\setlength\itemsep{-0.2em}
\item[(i)] Model ST1$_{1}$ if $\phi_2=0$, i.e.  area-level temporal Poisson mixed model with SAR(1) spatial domain effects;
\item[(ii)] Model T1 if $\rho=0$, i.e. area-level temporal Poisson mixed model with independent domain effects and independent domain-time effects;
\item[(iii)] Model T1$_{2}$ if $\phi_1=0$, i.e. area-level temporal Poisson mixed model with independent domain-time effects;
\item[(iv)] Model S1, if $\phi_2=0$ and $T=1$, i.e. area-level Poisson mixed model with SAR(1) spatial domain effects;
\item[(v)] Model 1 if $\rho=\phi_2=0$ and $T=1$, i.e.  area-level Poisson mixed model with independent domain effects;
\item[(vi)] Model 0 if $\rho=\phi_1=\phi_2=0$ and $T=1$, i.e. area-level Poisson regression model.
\end{itemize}
Boubeta et al. (2016, 2017) gives applications of Models 1, T1$_{2}$ and T1 to SAE problems.
This paper presents statistical methodology for Model ST1.
The corresponding procedures (fitting algorithms, predictors or MSE estimators) for the above cited submodels can be obtained by straightforward particularizations.
More general models, as area-level Poisson mixed model with SAR(1) spatial domain effects and AR(1) correlated time effects (Model ST2),
are not considered because the number of time instants $T=14$ and $T=1$ are rather small in the applications to fire data and poverty, respectively.
In addition, the first application considers the data of fires for two consecutive years in the months of April to October.
That is the period in which forest fires occur in Galicia.
Therefore, the 14 periods are divided into two groups of 7 in which an autoregressive or moving average periodic modeling could be proposed.
This poses an excessive complexity to model data with few temporary instants.
For these reasons, we have chosen to only incorporate the correlation that comes from the auxiliary variables and
the information from the past of the objective and auxiliary variables.
That is, we limit the complexity of the possible models to take into account in Model ST1.

Under Model ST1, it holds that
\vspace*{-.3cm}
\begin{equation*}
\mathbb{P}(y_{dt}|\vv)= \mathbb{P}(y_{dt}|v_{dt})=\frac{1}{y_{dt}!}\exp\{-\nu_{dt}p_{dt}\}\nu_{dt}^{y_{dt}}p_{dt}^{y_{dt}},
\end{equation*}
where $p_{dt}=\exp\{\xx_{dt}\bbeta+\phi_1 v_{1,d}+\phi_2 v_{2,dt}\}$ represents the target parameter.
The probability function of the response variable
$\yy=\underset{1\leq d\leq D}{\hbox{col}}(\underset{1\leq t\leq T}{\hbox{col}}(y_{dt}))$, conditionally on the random effects $\vv = (\vv_1, \vv_2)$, is
\begin{equation*}
\mathbb{P}(\yy|\vv)=\prod_{d=1}^D\prod_{t=1}^T \mathbb{P}(y_{dt}|\vv).
\end{equation*}
The marginal probability function of  $\yy$ is
\begin{equation*}
\mathbb{P}(\yy)=\int_{\R^{D(T+1)}} \mathbb{P}(\yy|\vv) f_{v}(\vv_1,\vv_2)\,d\vv_1d\vv_2
=\int_{\R^{D(T+1)}} \psi(\yy,\vv)\,d\vv,
\end{equation*}
where
\begin{align*}
\psi(\yy,\vv)&=
f_v(\vv_1,\vv_2) \prod_{d=1}^D\prod_{t=1}^T\frac{\exp\{-\nu_{dt}p_{dt}\} \nu_{dt}^{y_{dt}}
p_{dt}^{y_{dt}}}{y_{dt}!}
\\
&=
c(\yy)|\GGamma(\rho)|^{-1/2}\exp\left\{-\frac12\,\vv_1^\prime\GGamma^{-1}(\rho)\vv_1-\frac12\,\vv_2^\prime\vv_2\right\}
\\
&\cdot
\exp\left\{-\sum_{d=1}^D\sum_{t=1}^T\nu_{dt}\exp\{\xx_{dt}\bbeta+ \phi_1 v_{1,d}+\phi_2 v_{2,dt}\}\right\}
\\
&\cdot
\exp\left\{\sum_{k=1}^p\big(\sum_{d=1}^D\sum_{t=1}^Ty_{dt}x_{dtk}\big) \beta_k +\phi_1\sum_{d=1}^Dy_{d.}v_{1,d}
+\phi_2\sum_{d=1}^D\sum_{t=1}^Ty_{dt}v_{2,dt}
\right\},
\end{align*}
with
$c(\yy)=(2\pi)^{-\frac{D(T+1)}{2}}\prod_{d=1}^D\prod_{t=1}^T\big(\nu_{dt}^{y_{dt}}/y_{dt}!\big)$
and $y_{d.}=\sum_{t=1}^Ty_{dt}$.
\medskip
%
%
%
%
%

The method of moments (MM) can be used to estimate the vector of parameters, $\ttheta=(\bbeta^\prime,\phi_1,\phi_2,\rho)^\prime$, of Model ST1.
A set of natural equations for applying the MM algorithm is
\begin{eqnarray}\label{secMM-MM}
\begin{split}
0&=f_k(\ttheta)=\frac{1}{DT}\sum_{d=1}^D\sum_{t=1}^T \mathbb{E}_{\ttheta}[y_{dt}]x_{dtk}
-\frac{1}{DT}\sum_{d=1}^D\sum_{t=1}^Ty_{dt}x_{dtk},\quad k=1,\ldots, p,
\\
0&=f_{p+1}(\ttheta)=\frac{1}{D}\sum_{d=1}^D \mathbb{E}_{\ttheta}[y_{d.}^2]-\frac{1}{D}\sum_{d=1}^Dy_{d.}^2,
\\
0&=f_{p+2}(\ttheta)=\frac{1}{DT}\sum_{d=1}^D\sum_{t=1}^T \mathbb{E}_{\ttheta}[y_{dt}^2]
-\frac{1}{DT}\sum_{d=1}^D\sum_{t=1}^Ty_{dt}^2,
\\
0&=f_{p+3}(\ttheta)=\frac{1}{D(D-1)}\sum_{d_1\neq d_2}^D \mathbb{E}_{\ttheta}[y_{d_1.}y_{d_2.}]
-\frac{1}{D(D-1)}\sum_{d_1\neq d_2}^Dy_{d_1.}y_{d_2.},
\end{split}
\end{eqnarray}
The MM estimator of $\ttheta$ is obtained by solving the system (\ref{secMM-MM}) of nonlinear equations. The updating formula of the Newton-Raphson algorithm is
\begin{equation}\label{sec3.1-NR}
\ttheta^{(m+1)}=\ttheta^{(m)}-\HH^{-1}(\ttheta^{(m)})\ff(\ttheta^{(m)}),
\end{equation}
where
\begin{equation}\label{componentsMMST1}
\ttheta=\underset{1\leq k \leq p+3}{\hbox{col}}(\theta_k),\quad
\ff(\ttheta)=\underset{1\leq k \leq p+3}{\hbox{col}}(f_k(\ttheta)),\quad
\HH(\ttheta)=\left(\frac{\partial f_k(\ttheta)}{\partial \theta_\ell}\right)_{k,\ell=1,\ldots,p+3}.
\end{equation}
Appendix A contains the calculations of the expectations appearing in $\ff(\ttheta)$ and $\HH(\ttheta)$.
The MM algorithm under Model ST1 keeps the steps of Algorithm (\ref{sec3.1-NR}), replacing $\ttheta$, $\HH$ and $\ff$ for those given in (\ref{componentsMMST1}).
As algorithm seeds for $\bbeta$, $\phi_1$ and $\phi_2$, the algorithm may take the ML estimators under the model with no spatial correlation ($\rho=0$), i.e. under Model T1.
They can be obtained by using, for example, the function \texttt{glmer} of the R library \texttt{lme4}.
For $\rho$, it may take the Moran's I measure of spatial autocorrelation
\begin{equation}\label{MoranI}
I=\frac{D}{\sum_{d_1=1}^D\sum_{d_2=1}^Dw_{d_1d_2}}
\frac{\sum_{d_1=1}^D\sum_{d_2=1}^Dw_{d_1d_2}(\tilde{v}_{1,d_1}-\tilde{v}_{1})(\tilde{v}_{1,d_2}-\tilde{v}_{1})}
{\sum_{d=1}^D(\tilde{v}_{1,d}-\tilde{v})^2},
\end{equation}
where $\tilde{v}_{1,d}$, $d=1,\ldots, D$, are the predicted random effects under Model T1, $\tilde{v}=\frac{1}{D}\sum_{d=1}^D\tilde{v}_{1,d}$ and the $w_{d_1d_2}$'s are the elements of the proximity matrix $\WW$.
The asymptotic variance of the MM estimator under Model ST1 can be approximated by a similar bootstrap algorithm to that described by Boubeta et al. (2017).

%
%
This section ends by presenting three bootstrap algorithms for testing the significance of the variance parameters, $\phi_1$ and $\phi_2$, and of the autocorrelation parameter $\rho$. Algorithm \ref{BootstrapTestPhi1MST1} gives a bootstrap procedure to test the hypothesis $H_0:\, \phi_1 = 0$. We test Model T1$_2$ against Model ST1.
\begin{algorithm}[H]
\begin{algorithmic}[1]
\caption{A bootstrap test for $H_0: \phi_1 = 0$}
\label{BootstrapTestPhi1MST1}
\State Fit the Model ST1 to data and calculate $\hat{\bbeta}$, $\hat{\phi}_1$, $\hat{\phi}_2$ and $\hat{\rho}$.
\State Fit the Model T$1_2$ to data and calculate $\hat{\bbeta}^0$ and $\hat{\phi}_2^0$.
\State For $b=1,\ldots, B$, do
\begin{itemize}
\item[i)] Generate a bootstrap resample under $H_0: \phi_1 = 0$, i.e.
\vspace*{-.2cm}
\begin{align*}
v_{2, dt}^{*(b)} &\sim N(0,1),\,\, p_{dt}^{*(b)}=\exp \{ \xx_{dt} \hat{\bbeta}^0 + \hat{\phi}_2^0 v_{2,dt}^{*(b)} \},
\\
y_{dt}^{*(b)} &\sim \mbox{Poiss} ( \nu_{dt} p_{dt}^{*(b)}), \,\, d=1,\ldots, D, \, t=1,\ldots, T.
\end{align*}
\item[ii)]\vspace*{-.2cm} Fit the Model ST1 to the bootstrap data $(y_{dt}^{*(b)}, \xx_{dt}), \, d=1, \ldots, D, \, t=1, \ldots, T$, and calculate $\hat{\bbeta}^{*(b)}$, $\hat{\phi}_1^{*(b)}$, $\hat{\phi}_2^{*(b)}$ and $\hat{\rho}^{*(b)}$.
\end{itemize}
\State Calculate the $p$-value:
$p= B^{-1} \#\{ \hat{\phi}_1^{*(b)} > \hat{\phi}_1 \}$.
\end{algorithmic}
\end{algorithm}
If the null hypothesis $H_0:\phi_1=0$ is rejected, the significance of the autocorrelation parameter can be tested.
Algorithm \ref{BootstrapTestRhoMST1} presents a bootstrap procedure for testing $H_0:\rho=0$.
We test Model T1 against Model ST1.
\begin{algorithm}[H]
\begin{algorithmic}[1]
\caption{A bootstrap test for $H_0: \rho = 0$}
\label{BootstrapTestRhoMST1}
\State Fit the Model ST1 to data and calculate $\hat{\bbeta}$, $\hat{\phi}_1$, $\hat{\phi}_2$ and $\hat{\rho}$.
\State Fit the Model T1 to data and calculate $\hat{\bbeta}^0$, $\hat{\phi}_1^0$ and $\hat{\phi}_2^0$.
\State For $b=1,\ldots, B$, do
\begin{itemize}
\item[i)] Generate a bootstrap resample under $H_0: \rho = 0$, i.e.
\vspace*{-.2cm}
\begin{align*}
v_{1, d}^{*(b)} &\sim N(0,1), \, v_{2, dt}^{*(b)} \sim N(0,1),\,\, p_{dt}^{*(b)}=\exp \{ \xx_{dt} \hat{\bbeta}^0 + \hat{\phi}_1^0 v_{1,d}^{*(b)} + \hat{\phi}_2^0 v_{2,dt}^{*(b)} \},
\\
y_{dt}^{*(b)} &\sim \mbox{Poiss} ( \nu_{dt} p_{dt}^{*(b)}), \,\, d=1,\ldots, D, \, t=1,\ldots, T.
\end{align*}
\item[ii)]\vspace*{-.2cm} Fit the Model ST1 to the bootstrap data $(y_{dt}^{*(b)}, \xx_{dt}), \, d=1, \ldots, D, \, t=1, \ldots, T$, and calculate $\hat{\bbeta}^{*(b)}$, $\hat{\phi}_1^{*(b)}$, $\hat{\phi}_2^{*(b)}$ and $\hat{\rho}^{*(b)}$.
\end{itemize}
\State Calculate the $p$-value:
$p=B^{-1} \#\{ \abs{\hat{\rho}^{*(b)}} > \abs{\hat{\rho}} \}$.
\end{algorithmic}
\end{algorithm}
Finally, Algorithm \ref{BootstrapTestPhi2MST1} gives a bootstrap procedure for testing the null hypothesis $H_0: \phi_2 = 0$.
We test Model ST1$_1$ against Model ST1.
\begin{algorithm}[H]
\begin{algorithmic}[1]
\caption{A bootstrap test for $H_0: \phi_2 = 0$}
\label{BootstrapTestPhi2MST1}
\State Fit the Model ST1 to data and calculate $\hat{\bbeta}$, $\hat{\phi}_1$, $\hat{\phi}_2$ and $\hat{\rho}$.
\State Fit the Model ST1$_1$ to data and calculate $\hat{\bbeta}^0$, $\hat{\phi}_1^0$ and $\hat{\rho}^0$.
\State For $b=1,\ldots, B$, do
\begin{itemize}
\item[i)] Generate a bootstrap resample under $H_0: \phi_2 = 0$, i.e.
\vspace*{-.2cm}
\begin{align*}
\vv_{1}^{*(b)} &\sim N_D(\cero,\GGamma (\hat{\rho}^0)),\,\, p_{dt}^{*(b)}=\exp \{ \xx_{dt} \hat{\bbeta}^0 + \hat{\phi}_1^0 v_{1,d}^{*(b)} \},
\\
y_{dt}^{*(b)} &\sim \mbox{Poiss} ( \nu_{dt} p_{dt}^{*(b)}), \,\, d=1,\ldots, D, \, t=1,\ldots, T.
\end{align*}
\item[ii)]\vspace*{-.2cm} Fit the Model ST1 to the bootstrap data $(y_{dt}^{*(b)}, \xx_{dt}), \, d=1, \ldots, D, \, t=1, \ldots, T$, and calculate $\hat{\bbeta}^{*(b)}$, $\hat{\phi}_1^{*(b)}$, $\hat{\phi}_2^{*(b)}$ and $\hat{\rho}^{*(b)}$.
\end{itemize}
\State Calculate the $p$-value:
$p= B^{-1}\#\{ \hat{\phi}_2^{*(b)} > \hat{\phi}_2 \}$.
\end{algorithmic}
\end{algorithm}
%
%
%
%
\section{The predictors}\label{sec3}
%
This section gives the EBP and a plug-in predictor of $p_{dt}$ under Model ST1. The EBP of $p_{dt}$ is obtained from the corresponding  best predictor (BP) by replacing the vector of model parameters $\ttheta$ by an estimator $\hat{\ttheta}$.
As the MM estimators are consistent, they are employed for calculating the EBPs.
To avoid overflow numerical problems in the calculation of the exact EBP, this section proposes alternative approximations.

Let $\yy_{d}$ be the response vector within the domain $d$, i.e. $\yy_{d}=\underset{1\leq t \leq T}{\mbox{col}}(y_{dt})$.
The conditional distribution of the response variable $\yy$, given the random effects $\vv_1$ and $\vv_2$, is
\begin{equation*}
\mathbb{P}(\yy|\vv_{1},\vv_{2})=\prod_{d=1}^D \mathbb{P}(\yy_d|v_{1,d},\vv_{2,d}),\quad
\mathbb{P}(\yy_d|v_{1,d},\vv_{2,d})=\prod_{t=1}^T \mathbb{P}(y_{dt}|v_{1,d},v_{2,dt}),
\end{equation*}
where
\begin{align*}
\mathbb{P}(y_{dt}|v_{1,d},v_{2,dt})&=\frac{1}{y_{dt}!}\exp\{-\nu_{dt}p_{dt}\}\nu_{dt}^{y_{dt}}p_{dt}^{y_{dt}}
\\
&=c_{dt}
\exp\left\{y_{dt}(\xx_{dt}\bbeta+\phi_1 v_{1,d}+\phi_2 v_{2,dt})
-\nu_{dt}\exp\{\xx_{dt}\bbeta+\phi_1 v_{1,d}+\phi_2 v_{2,dt}\}\right\}.
\end{align*}
%
%
\subsection{The empirical best predictor}\label{sec3.1}
%
%
The BP of $p_{dt}$ is the unbiased predictor minimizing the MSE. It is obtained as the conditional expectation $\hat{p}_{dt}(\ttheta)= \mathbb{E}_{\ttheta}[p_{dt}|\yy]$. It holds that
\begin{equation}\label{EBPModelST1}
\mathbb{E}_{\ttheta}[p_{dt}|\yy]=
\frac{\int_{\R^{D(T+1)}}p_{dt} \mathbb{P} ( \yy \vert \vv_1, \vv_2 ) f(\vv_1) f(\vv_2)\,d\vv_{1}d\vv_{2}}
{\int_{\R^{D(T+1)}}\mathbb{P} ( \yy \vert \vv_1, \vv_2 ) f(\vv_1) f(\vv_2)\,d\vv_{1}d\vv_{2}}=\frac{N_{dt}(\yy,\ttheta)}{B(\yy,\ttheta)},
\end{equation}
where
\begin{align}
N_{dt}(\yy,\ttheta)&=
\int_{\R^{D(T+1)}}\exp\{\xx_{dt}\bbeta+\phi_1v_{1,d}+\phi_2v_{2,dt}\}
( \prod_{\ell=1}^D\prod_{\tau=1}^T \mathbb{P}(y_{\ell\tau}|v_{1,\ell},v_{2,\ell\tau}) ) f(\vv_{1})f(\vv_{2})\,d\vv_{1}d\vv_{2}\nonumber
\\
&=
\int_{\R^{D(T+1)}}\prod_{\ell=1}^D\prod_{\tau=1}^{T}
\exp\Big\{(y_{\ell\tau}+\delta_{d\ell}\delta_{t\tau})(\xx_{\ell\tau}\bbeta+\phi_1v_{1,\ell}+\phi_2v_{2,d\tau})\nonumber
\\
&-
\nu_{\ell\tau}\exp\{\xx_{\ell\tau}\bbeta+\phi_1v_{1,\ell}+\phi_2v_{2,\ell\tau}\}\Big\}
f(\vv_{1})f(\vv_{2})\,d\vv_{1}d\vv_{2},\nonumber
\end{align}
\begin{align}\label{BModelST1}
B(\yy,\ttheta) &=
\int_{\R^{D(T+1)}}
( \prod_{\ell=1}^D\prod_{\tau=1}^T \mathbb{P}(y_{\ell\tau}|v_{1,\ell},v_{2,\ell \tau})) f(\vv_{1})f(\vv_{2})\,d\vv_{1}d\vv_{2}\nonumber
\\
&=
\int_{\R^{D(T+1)}}\prod_{\ell=1}^D\prod_{\tau=1}^{T}
\exp\Big\{y_{\ell\tau}(\xx_{\ell\tau}\bbeta+\phi_1v_{1,\ell}+\phi_2 v_{2,d\tau})\nonumber
\\
&-
\nu_{\ell\tau}\exp\{\xx_{\ell\tau}\bbeta+\phi_1v_{1,\ell}+\phi_2v_{2,\ell\tau}\}\Big\}
f(\vv_{1})f(\vv_{2})\,d\vv_{1}d\vv_{2},
\end{align}
and $\delta_{d\ell}$ and $\delta_{t\tau}$ are Kronecker deltas, i.e. $\delta_{ij}=1$ if $i=j$ and $\delta_{ij}=0$ otherwise.
The numerator $N_{dt}(\yy,\ttheta)$ can be expressed in terms of the denominator $B(\yy,\ttheta)$ as $N_{dt}(\yy,\ttheta) = B(\yy + \ee_{dt},\ttheta)$,
where
$\ee_{dt}=\underset{1\leq \ell\leq D}{\hbox{col}}(\underset{1\leq \tau\leq T}{\hbox{col}}(\delta_{d\ell}\delta_{t\tau}))$.
The EBP of $p_{dt}$ is $\hat{p}_{dt} = \hat{p}_{dt}(\hat{\ttheta})$.

The EBP calculation involves complex integrals in a high-dimensional space.
The integrals are approximated by using an antithetic Monte Carlo algorithm. The steps are:
\begin{enumerate}
\item
Generate $\vv_{1}^{(s_1)}\sim N_D\big(\cero,\GGamma(\hat{\rho})\big)$, $v_{2,\ell\tau}^{(s_2)}$ i.i.d. $N(0,1)$ and calculate
$\vv_{1}^{(S_1+s_1)}=-\vv_{1}^{(s_1)}$, $v_{2,\ell\tau}^{(S_2+s_2)}=-v_{2,\ell\tau}^{(s_2)}$, $s_1=1,\ldots,S_1$, $s_2=1,\ldots,S_2$, $\ell=1,\ldots,D$, $\tau=1,\ldots,T$.
\item
Approximate the EBP of $p_{dt}$ by $\hat{p}_{dt}(\hat\ttheta)=\hat{N}_{dt}(\yy,\hat\ttheta)/\hat{B}(\yy,\hat\ttheta)$, where \begin{align}\label{sec2-BdHat1ModelST1}
\hat{B}(\yy,\hat\ttheta) &=\sum_{s_1=1}^{2S_1}\sum_{s_2=1}^{2S_2}\prod_{\ell=1}^D\prod_{\tau=1}^{T}\exp\left\{y_{\ell\tau}
(\xx_{\ell\tau}\hat\bbeta+\hat\phi_1v_{1,\ell}^{(s_1)}+\hat\phi_2v_{2,\ell\tau}^{(s_2)})\right. \nonumber
\\
&-\left.
\nu_{\ell\tau}\exp\{\xx_{\ell\tau}\hat\bbeta+\hat\phi_1v_{1,\ell}^{(s_1)}+\hat\phi_2v_{2,\ell\tau}^{(s_2)}\}\right\},\quad
\hat{N}_{dt}(\yy,\hat\ttheta)=\hat{B}(\yy + \ee_{dt},\hat\ttheta).
\end{align}
\end{enumerate}
As the relationship between the mean and probability parameters is $\mu_{dt}=\nu_{dt} p_{dt}$, and $\nu_{dt}$ is known (size parameter), the EBP of $\mu_{dt}$ is obtained as an immediate consequence of the EBP of $p_{dt}$. That is to say, the EBP of $\mu_{dt}$ is $\hat\mu_{dt}(\hat\ttheta)=\nu_{dt}\hat{p}_{dt}(\hat\ttheta)$.

The EBP calculations in (\ref{sec2-BdHat1ModelST1}) are computationally demanding. For this reason, we propose an approximation to the BP of $p_{dt}$ (\ref{EBPModelST1}) under Model ST1.
Divide $\yy$ and $\vv=(\vv_1^\prime,\vv_2^\prime)^\prime$ into two parts $(\yy_d^\prime,\yy_{d-}^\prime)^\prime$ and $(\vv_d^\prime,\vv_{d-}^\prime)^\prime$, where
$\yy_d=\underset{1\leq t\leq T}{\hbox{col}}(y_{dt})$
$\yy_{d-}=\underset{1\leq i\leq D,\, i\neq d}{\hbox{col}}(\yy_i)$,
$\vv_d=(v_{1,d},\vv_{2,d}^\prime)^\prime$ and
$\vv_{d-}=\underset{1\leq i\leq D,\, i\neq d}{\hbox{col}}(\vv_i)$.
The conditional distribution of $\yy$, given $\vv$, is
\begin{equation}\label{chap3:sec3.1-condProbST1}
\mathbb{P}(\yy|\vv)=\prod_{i=1}^D \mathbb{P}(\yy_i|\vv_i) = \mathbb{P}(\yy_d|\vv_d)\prod_{i=1, i\neq d}^D \mathbb{P}(\yy_i|\vv_i)= \mathbb{P}(\yy_d|\vv_d) \mathbb{P}(\yy_{d-}|\vv_{d-}).
\end{equation}
The p.d.f. of $\vv_d$ is
\begin{equation*}
f(\vv_d)=f(v_{1,d})f(\vv_{2,d}),
\end{equation*}
where $v_{1,d}\sim N(0, \gamma_{dd}(\rho))$ and $\vv_{2,d}\sim N(\cero,\II_T)$.
The component $B(\yy,\ttheta)$ of the BP of $p_{dt}$ in (\ref{EBPModelST1}) can be rewritten by using the decomposition of the conditional probability given in (\ref{chap3:sec3.1-condProbST1}), i.e.
\begin{align*}
B(\yy,\ttheta) =
\int_{\R^{T+1}}
\Big[\int_{\R^{(D-1)(T+1)}} \mathbb{P}(\yy_{d-}|\vv_{d-})f(\vv_{d-}|\vv_d)\,d\vv_{d-}\Big] \mathbb{P}(\yy_d|\vv_d)f(\vv_d)\,d\vv_d.
\end{align*}
As $\mathbb{P}(\yy_{d-}|\vv_{d-})f(\vv_{d-}|\vv_d)= \mathbb{P}(\yy_{d-}|\vv_{d-},\vv_d)f(\vv_{d-}|\vv_d)$, the inner integral is
\begin{equation*}
\int_{\R^{(D-1)(T+1)}}\mathbb{P}(\yy_{d-}|\vv_{d-},\vv_d)f(\vv_{d-}|\vv_d)\,d\vv_{d-}= \mathbb{P}(\yy_{d-}|\vv_d).
\end{equation*}
Therefore, it holds that
\begin{align*}
B(\yy,\ttheta) &= \int_{\R^{T+1}} \mathbb{P}(\yy_{d-}|\vv_d)\mathbb{P}(\yy_d|\vv_d) f(\vv_d)\,d\vv_d.
\end{align*}
By applying similar developments as for the component $N_{dt}(\yy, \ttheta)$ of (\ref{EBPModelST1}),  it holds that
\begin{align*}
N_{dt}(\yy, \ttheta) &= \int_{\R^{T+1}}\exp\{\xx_{dt}\bbeta+\phi_1v_{1,d}+\phi_2 v_{2,dt}\} \mathbb{P}(\yy_{d-}|\vv_d)\mathbb{P}(\yy_d|\vv_d)f(\vv_d)\,d\vv_d.
\end{align*}
Under the assumption
\begin{equation}\label{ApproxMST1}
\PR(\yy_{d-}|\vv_d)\approx \PR(\yy_{d-}),
\end{equation}
the BP of $p_{dt}$, $\hat{p}_{dt}(\ttheta)$, can be approximated by
\begin{equation*}
\hat{p}_{dt}^a(\ttheta) = N_{dt}^a(\yy_d, \ttheta) / B_d^a(\yy_d, \ttheta),
\end{equation*}
where
\begin{align}\label{BdAppoxMST1}
N_{dt}^a(\yy_d, \ttheta) &=
\int_{\R^{T+1}}\exp\{\xx_{dt}\bbeta+\phi_1v_{1,d}+\phi_2 v_{2,dt}\}
\prod_{\tau=1}^T \mathbb{P}(y_{d\tau}|v_{1,d},v_{2,d\tau})f(v_{1,d})f(\vv_{2,d})\,dv_{1,d}\,d\vv_{2,d}\nonumber
\\
&= B_d^a(\yy_d+\eepsilon_t, \ttheta),\quad \eepsilon_t=\underset{1\leq \tau\leq T}{\hbox{col}}(\delta_{t\tau}),\nonumber
\\
B_d^a(\yy_d, \ttheta) &=
\int_\R \prod_{\tau=1}^T\Big[\int_\R \exp\Big\{y_{d\tau}(\xx_{d\tau}\bbeta+\phi_1v_{1,d}+\phi_2 v_{2,d\tau})\nonumber
\\
&-
\exp\{\xx_{d\tau}\bbeta+\phi_1v_{1,d}+\phi_2 v_{2,d\tau}\}\Big\}
f(v_{2,d\tau})dv_{2,d\tau}\Big]f(v_{1,d})dv_{1,d}.
\end{align}
Then, the EBP of $p_{dt}$, $\hat{p}_{dt}(\hat\ttheta)$, can be approximated as follows.
\begin{enumerate}
\item
Estimate $\hat{\ttheta}=(\hat\bbeta^\prime,\hat{\phi}_1,\hat{\phi}_2,\hat{\rho})$.
\item
For $s_1=1,\ldots,S_1$, $s_2=1,\ldots,S_2$, $\ell=1,\ldots,D$, $\tau=1,\ldots,T$, generate $\vv_{1}^{(s_1)}\sim N_D\big(\cero,\GGamma(\hat{\rho})\big)$, $v_{2,\ell\tau}^{(s_2)}$ i.i.d. $N(0,1)$ and calculate $\vv_{1}^{(S_1+s_1)}=-\vv_{1}^{(s_1)}$, $v_{2,\ell\tau}^{(S_2+s_2)}=-v_{2,\ell\tau}^{(s_2)}$.
\item
Calculate $\hat{p}_{dt}^a(\hat\ttheta)=\hat{N}_{dt}^a(\yy_d,\hat\ttheta)/\hat{B}_d^a(\yy_d,\hat\ttheta)$, $d=1,\ldots,D$, $t=1,\ldots,T$, where 
\begin{align*}
\hat{B}_d^a(\yy_d,\hat\ttheta) &= \sum_{s_1=1}^{2S_1}\prod_{\tau=1}^{T}\sum_{s_2=1}^{2S_2}\exp\left\{y_{d\tau}
(\xx_{d\tau}\hat\bbeta+\hat\phi_1v_{1,d}^{(s_1)}+\hat\phi_2v_{2,d\tau}^{(s_2)})\right.
\\
&-\left.
\nu_{d\tau}\exp\{\xx_{d\tau}\hat\bbeta+\hat\phi_1v_{1,d}^{(s_1)}+\hat\phi_2v_{2,d\tau}^{(s_2)}\}\right\},\quad
\hat{N}_{dt}^a(\yy_d,\hat\ttheta)=B_d^a(\yy_d+\eepsilon_t,\hat\ttheta).
\end{align*}
\end{enumerate}
The EBP approximation, $\hat{p}_{dt}^a(\hat{\ttheta})$, maintains the expression of the EBP of $p_{dt}$ under the area-level Poisson mixed model with independent time effects studied by Boubeta et al (2017). However, the domain random effects, $v_{1, d}$, are now generated according to a SAR(1) process.
%
%
%
\subsection{The plug-in predictor}\label{sec3.2}
%
The plug-in predictor of $p_{dt}$ is obtained by replacing, in the theoretical expression of $p_{dt}$, the model parameters by their estimates and the random effects by their EBPs, i.e.
\begin{equation}\label{plugMST1}
\hat{p}_{dt}^{P}=\exp\{ \xx_{dt}\hat\bbeta+\hat\phi_1\hat{v}_{1,d}+\hat\phi_2\hat{v}_{2,dt}\}.
\end{equation}
As above, the EBPs of $\vv_1$ and $\vv_2$ are obtained from the respective BPs.
The BP of $v_{1,d}$ is
\begin{equation*}
\hat{v}_{1,d}(\ttheta)=\mathbb{E}_{\ttheta}[v_{1,d}|\yy]=\frac{N_{1,d}(\yy,\ttheta)}{B(\yy,\ttheta)},
\end{equation*}
where $B(\yy,\ttheta)$ was defined in (\ref{BModelST1}) and
\begin{align*}
N_{1,d}(\yy,\ttheta)&=
\int_{\R^{D(T+1)}}v_{1,d}
(\prod_{\ell=1}^D\prod_{\tau=1}^T \mathbb{P}(y_{\ell\tau}|v_{1,\ell},\vv_{2,\ell})) f(\vv_{1})f(\vv_{2})\,d\vv_{1}d\vv_{2}
\\
&=
\int_{\R^{D(T+1)}}\prod_{\ell=1}^D\prod_{\tau=1}^{T}
\exp\Big\{y_{\ell\tau}(\xx_{\ell\tau}\bbeta+\phi_1v_{1,\ell}+\phi_2v_{2,\ell\tau})
\\
&-
\nu_{\ell\tau}\exp\{\xx_{\ell\tau}\bbeta+\phi_1v_{1,\ell}+\phi_2v_{2,\ell\tau}\}\Big\}
v_{1,d}f(\vv_{1})f(\vv_{2})\,d\vv_{1}d\vv_{2}.
\end{align*}
The EBP of $v_{1,d}$ is $\hat{v}_{1,d}=\hat{v}_{1,d}(\hat\ttheta)$ and it can be approximated as follows.
\begin{enumerate}
\item
Generate $\vv_{1}^{(s_1)}\sim N_D\big(\cero,\GGamma(\hat{\rho})\big)$, $v_{2,\ell\tau}^{(s_2)}$ i.i.d. $N(0,1)$ and calculate
$\vv_{1}^{(S_1+s_1)}=-\vv_{1}^{(s_1)}$, $v_{2,\ell\tau}^{(S_2+s_2)}=-v_{2,\ell\tau}^{(s_2)}$, $s_1=1,\ldots,S_1$, $s_2=1,\ldots,S_2$, $\ell=1,\ldots,D$, $\tau=1,\ldots,T$.
\item
Calculate $\hat{v}_{1,d}(\hat\ttheta)=\hat{N}_{1,d}(\yy,\hat\ttheta)/\hat{B}(\yy,\hat\ttheta)$, where $\hat{B}(\yy,\hat\ttheta)$ was defined in (\ref{sec2-BdHat1ModelST1}) and
\begin{align*}
\hat{N}_{1,d}(\yy,\hat\ttheta)&=\sum_{s_1=1}^{2S_1}\sum_{s_2=1}^{2S_2}\prod_{\ell=1}^D\big\{1+\delta_{\ell d}(v_{1,\ell}^{(s_1)}-1)\big\}\prod_{\tau=1}^{T}
\exp\left\{y_{\ell\tau}
(\xx_{\ell\tau}\hat\bbeta+\hat\phi_1v_{1,\ell}^{(s_1)}+\hat\phi_2v_{2,\ell\tau}^{(s_2)})\right.
\\
&-\left.
\nu_{\ell\tau}\exp\{\xx_{\ell\tau}\hat\bbeta+\hat\phi_1v_{1,\ell}^{(s_1)}+\hat\phi_2v_{2,\ell\tau}^{(s_2)}\}\right\}.
\end{align*}
\end{enumerate}
The BP of the domain-time random effects $v_{2,dt}$ is
\begin{equation*}
\hat{v}_{2,dt}(\ttheta)=\mathbb{E}_{\ttheta}[v_{2,dt}|\yy]=\frac{N_{2,dt}(\yy,\ttheta)}{B(\yy,\ttheta)},
\end{equation*}
where
 $B(\yy,\ttheta)$ was defined in (\ref{BModelST1}) and
\begin{align*}
N_{2,dt}(\yy,\ttheta)&=
\int_{\R^{D(T+1)}}v_{2,dt}
\big(\prod_{\ell=1}^D\prod_{\tau=1}^T \mathbb{P}(y_{\ell\tau}|v_{1,\ell},\vv_{2,\ell})\Big)f(\vv_{1})f(\vv_{2})\,d\vv_{1}d\vv_{2}
\\
&=
\int_{\R^{D(T+1)}}\prod_{\ell=1}^D\prod_{\tau=1}^{T}
\exp\Big\{y_{\ell\tau}(\xx_{\ell\tau}\bbeta+\phi_1v_{1,\ell}+\phi_2v_{2,d\tau})
\\
&-
\nu_{\ell\tau}\exp\{\xx_{\ell\tau}\bbeta+\phi_1v_{1,\ell}+\phi_2v_{2,\ell\tau}\}\Big\}
v_{2,dt}f(\vv_{1})f(\vv_{2})\,d\vv_{1}d\vv_{2}.
\end{align*}
The EBP of $v_{2,dt}$ is $\hat{v}_{2,dt}=\hat{v}_{2,dt}(\hat\ttheta)$ and it can be approximated as follows.
\begin{enumerate}
\item
Generate $\vv_{1}^{(s_1)}\sim N_D\big(\cero,\GGamma(\hat{\rho})\big)$, $v_{2,\ell\tau}^{(s_2)}$ i.i.d. $N(0,1)$ and calculate
$\vv_{1}^{(S_1+s_1)}=-\vv_{1}^{(s_1)}$, $v_{2,\ell\tau}^{(S_2+s_2)}=-v_{2,\ell\tau}^{(s_2)}$, $s_1=1,\ldots,S_1$, $s_2=1,\ldots,S_2$, $\ell=1,\ldots,D$, $\tau=1,\ldots,T$.
\item
Calculate $\hat{v}_{2,dt}(\hat\ttheta)=\hat{N}_{2,dt}(\yy,\hat\ttheta)/\hat{B}(\yy,\hat\ttheta)$,  where $\hat{B}(\yy,\hat\ttheta)$ was defined in (\ref{sec2-BdHat1ModelST1}) and
\begin{align*}
\hat{N}_{2,dt}(\yy,\hat\ttheta)&=\sum_{s_1=1}^{2S_1}\sum_{s_2=1}^{2S_2}\prod_{\ell=1}^D\prod_{\tau=1}^{T}
\big[1+\delta_{d\ell}\delta_{t\tau}(v_{2,\ell\tau}^{(s_2)}-1)\big]
\exp\left\{y_{\ell\tau}
(\xx_{\ell\tau}\hat\bbeta+\hat\phi_1v_{1,\ell}^{(s_1)}+\hat\phi_2v_{2,\ell\tau}^{(s_2)})\right.
\\
&-\left.
\nu_{\ell\tau}\exp\{\xx_{\ell\tau}\hat\bbeta+\hat\phi_1v_{1,\ell}^{(s_1)}+\hat\phi_2v_{2,\ell\tau}^{(s_2)}\}\right\}.
\end{align*}
\end{enumerate}
The EBPs of $v_{1,d}$ and $v_{2,dt}$ are computationally demanding and therefore this section proposes approximations, $\hat{v}_{1,d}^a$ and $\hat{v}_{2,dt}^a$, to the EBPs $\hat{v}_{1,d}$ and $\hat{v}_{2,dt}$ respectively.
Under the assumption (\ref{ApproxMST1}), the BP of the domain random effects, $\hat{v}_{1,d}(\ttheta)$, can be approximated by
\begin{equation*}
\hat{v}_{1,d}^a(\ttheta) = \frac{N_{1,d}^a (\yy_{d},\ttheta)}{B_{d}^a (\yy_{d},\ttheta)},
\end{equation*}
where $B_{d}^a (\yy_{d},\ttheta)$ is given in (\ref{BdAppoxMST1}) and
\begin{eqnarray*}
N_{1,d}^a(\yy_{d},\ttheta)&=&
\int_{\R}\prod_{\tau=1}^{T}\bigg[\int_\R
\exp \Big\{ y_{d\tau}(\xx_{d\tau}\bbeta+\phi_1v_{1,d}+\phi_2v_{2,d\tau})
\\[-.15cm]
&-&
\nu_{d\tau}\exp\{\xx_{d\tau}\bbeta+\phi_1v_{1,d}+\phi_2v_{2,d\tau}\}\Big\}
f(v_{2,d\tau})\,dv_{2,d\tau}\bigg] v_{1,d}f(v_{1,d})\,dv_{1,d}.
\end{eqnarray*}
The difference between the approximated BP, $\hat{v}_{1, d}^a (\ttheta)$, and the BP based on Model T1 is that the distribution of the domain random effects is SAR(1) instead of being i.i.d. $N(0,1)$.

The BP of $v_{2,dt}$, $\hat{v}_{2,dt}(\ttheta)$, can be approximated by
\begin{equation*}
\hat{v}_{2,dt}^a(\ttheta) = \frac{N_{2,dt}^a (\yy_{d},\ttheta)}{B_{d}^a (\yy_{d},\ttheta)},
\end{equation*}
where
\begin{align*}
N_{2,dt}^a(\yy_{d},\ttheta) &=
\int_{\R}\prod_{\tau=1}^{T} \bigg[ \int_{\R} ( 1+\delta_{t\tau}(v_{2,d\tau}-1))
\exp \Big\{y_{d\tau}(\xx_{d\tau}\bbeta+\phi_1v_{1,d}+\phi_2v_{2,d\tau})
\\
&-
\nu_{d\tau}\exp\{\xx_{d\tau}\bbeta+\phi_1v_{1,d}+\phi_2v_{2,d\tau}\}\Big\} f(v_{2,d\tau})\,dv_{2,d\tau} \bigg] f(v_{1,d})\, dv_{1,d}.
\end{align*}
In this case, the underlying spatial correlation structure slightly complicates the expression of the approximated BP, $\hat{v}_{2,dt}^a(\ttheta)$, with respect to that obtained under Model T1.

To finish this section, we introduce  the synthetic predictors
\begin{equation}\label{synST1}
\hat{p}_{dt}^{syn}=\exp\{\xx_{dt}\hat\bbeta\},\quad \hat{\mu}_{dt}^{syn}=\nu_{dt}\hat{p}_{dt}^{syn},
\end{equation}
which do not require the calculation of the EBPs of the random effects.
The synthetic predictors can give a parsimonious solution in those cases in which the standard deviation parameters are very small and a set of highly informative auxiliary variables is available.
%
%
%
%
\subsection{MSE estimation}\label{sec3.3}
%
%
%
As accuracy measure of a predictor (EBP or plug-in) of $p_{dt}$ under Model ST1, this section considers the MSE.
It proposes estimating the MSE of the predictor of $p_{dt}$ by using a parametric bootstrap algorithm based on the bootstrap procedure given by Gonz\'alez-Manteiga et al. (2008, 2010), since the analytical estimation is not feasible computationally.
The steps are:
\begin{enumerate}
\item
Fit the model to the sample and calculate $\hat\ttheta=(\hat{\bbeta},\hat{\phi}_1,\hat{\phi}_2,\hat{\rho})$ under Model ST1.
\item
For $d=1,\ldots, D$, $t=1,\ldots, T$, repeat $B$ times, $b=1,\ldots,B$.
\begin{enumerate}
\item
Generate the bootstrap random effects $\vv_{1}^{*(b)}\sim N_D\big(\cero,\GGamma(\hat{\rho})\big)$ and $\{v_{2,dt}^{*(b)}\}$ i.i.d. $N(0,1)$.
\item
Calculate the theoretical bootstrap quantity $p_{dt}^{*(b)}=\text {exp}\{\xx_{dt}\hat\bbeta+\hat\phi_1 v_{1,d}^{*(b)}+\hat\phi_2 v_{2,dt}^{*(b)}\}$.
\item
Generate the response variables $y_{dt}^{*(b)}\sim\mbox{Poiss}(\nu_{dt}p_{dt}^{*(b)})$.
\item
Calculate  $\hat\ttheta^{*(b)}$ and the predictor (EBP or plug-in) $\hat{p}_{dt}^{*(b)}=\hat{p}_{dt}^{*(b)}(\hat\ttheta^{*(b)})$.
\end{enumerate}
\item
Output:
\begin{equation*}
mse^*(\hat{p}_{dt})=\frac{1}{B}\sum_{b=1}^B\big(\hat{p}_{dt}^{*(b)}-p_{dt}^{*(b)}\big)^2.
\end{equation*}
\end{enumerate}

%
%
\section{Application to fire data}\label{sec.firedata}
%
%
This section presents an application to real data of wildfires in Galicia during $2007-2008$.
The objective of this study is to analyse the target variable \textit{number of wildfires} by forest areas and months.
The domains are the forest areas. For each domain, the data set collects the number of wildfires by month.
The section only takes the months between April to October for being the months with the greatest number of fires.
This is the main reason for not considering models with time correlated random effects.
The total number of domains and time instants are $D=63$ and $T=14$, respectively.

The response variable at domain $d$, $d=1,\ldots, D$, and time $t$, $t=1,\ldots, T$, $y_{dt}$, is explained by some auxiliary variables through an
area-level spatio-temporal Poisson mixed model with SAR(1)-correlated domain and independent domain-time effects.
As proximity matrix, $\WW$, we take the common border option described in Section \ref{sec2}.
The auxiliary variables are \textit{acumRain} (accumulated rain in $l/m^2$), \textit{averTemp} (Average air temperature in $^{\circ}C$) and
\textit{cadHold} (Number of owners of cadastral parcels). The two first variables are calculated by averaging the measurements of meteorological stations by months and forests areas.
The third variable is annually updated in the Land Register of the Spanish Ministry of Agriculture, Fisheries and Food.
See Boubeta et al. (2019) for further information about the auxiliary variables.

Table \ref{APoissSAR1TindFit} presents the MM estimates of the fixed effects under Model ST1.
This table suggests that $acumRain$ is protective, since it causes a decrease in the response variable if it increases and the other variables remain fixed.
On the other hand, $averTemp$ and $cadHold$ are directly related to the response variable since their signs are positive.
The three covariates are significant at the level $\alpha=0.05$.
\renewcommand{\arraystretch}{1}
\begin{table}[H]
\centering
\caption{MM estimates of regression parameters under Model ST1.}
\label{APoissSAR1TindFit}
\begin{tabular}{lrrrr}
\toprule
Variable & Est. & s.e. & $z$-value &  $p$-value \\
\midrule
$Intercept$ & 0.7736 & 0.0870 & 8.8910 & $<$ 0.001 \\
$acumRain$  &-0.5207 & 0.0682 &-7.6318 & $<$ 0.001 \\
$averTemp$  & 0.1507 & 0.0661 & 2.2820 & 0.0225 \\
$cadHold$   & 0.2931 & 0.0605 & 4.8466 & $<$ 0.001 \\
\bottomrule
\end{tabular}
\end{table}

The MM estimates of the standard deviations are $\hat{\phi}_1=0.331$ and $\hat{\phi}_2=0.696$.
Their $95\%$ percentile bootstrap confidence intervals are $(0.130, 0.467)$ and $(0.503, 0.829)$ respectively.
The estimate of the spatial autocorrelation is $\hat{\rho}=0.327$.
The Algorithms \ref{BootstrapTestPhi1MST1} and \ref{BootstrapTestPhi2MST1} are used to test the significance of the parameters $\phi_1$ and $\phi_2$.
The obtained bootstrap $p$-values are $0.01$ and $0.00$ respectively.
The conclusion is that both standard deviations are significantly different from $0$.
In addition, as the hypothesis $H_0: \phi_1 = 0$ is rejected, the Algorithm \ref{BootstrapTestRhoMST1} is applied to test $H_0: \rho = 0$.
The obtained bootstrap $p$-value is $0.00$.
Then, based on the bootstrap $p$-values, 
 it is recommended to use a spatio-temporal Poisson mixed model ST1 to analyse wildfires in Galicia by forest areas and the considered months in the period  $2007-2008$.

\begin{figure}[h]
  \centering
  \begin{tabular}{cc}
  \includegraphics[width=6cm,height=6.4cm]{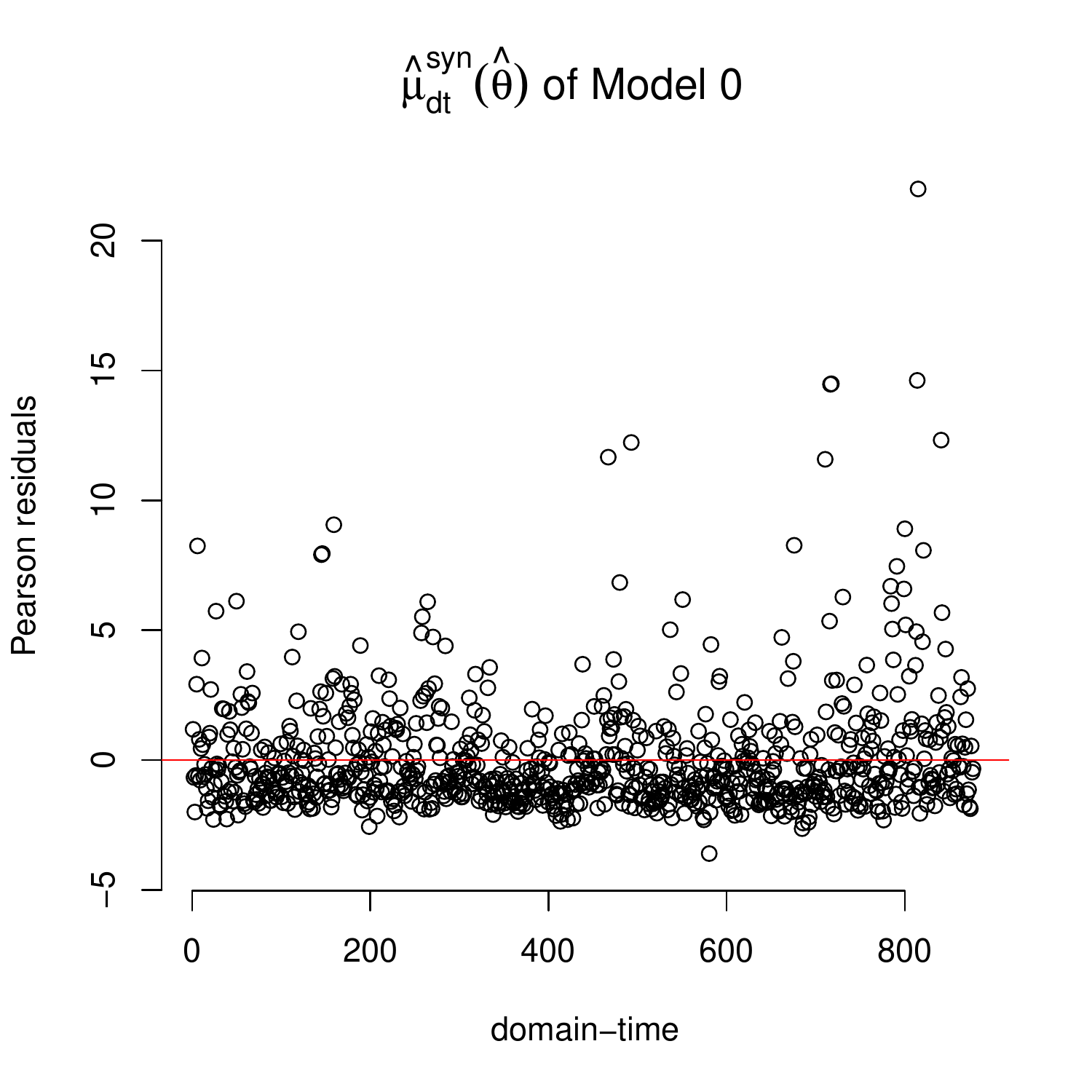} &
  \includegraphics[width=6cm,height=6.4cm]{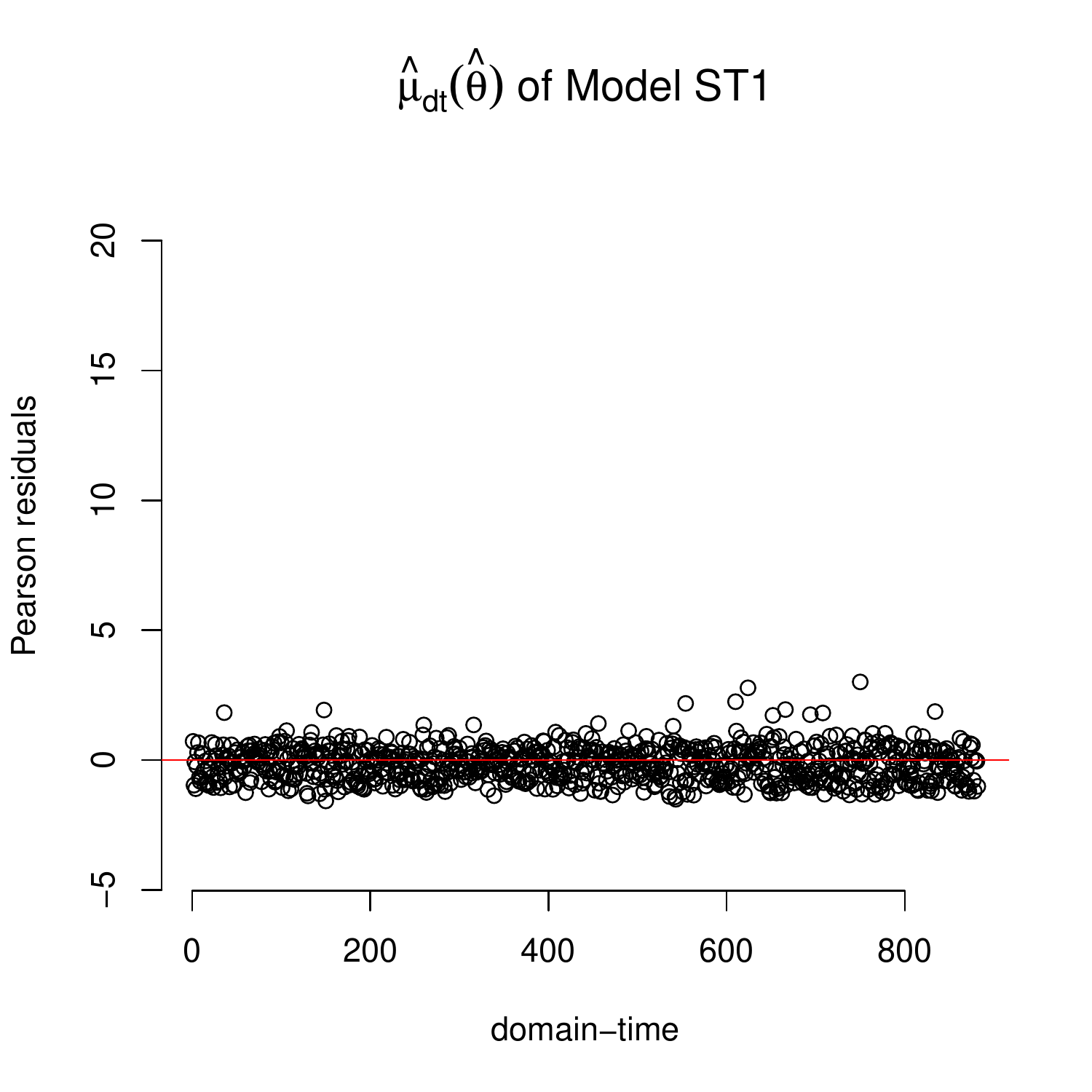}
  \end{tabular}
  \vspace*{-.2cm}
  \caption{Pearson residuals of the synthetic estimator under Model 0  (left) and of the EBP of $\mu_{dt}$ under Model ST1 (right).}
  \label{sec6-Res}
\end{figure}

Figure \ref{sec6-Res} plots the Pearson residuals of the synthetic estimator, $\hat{\mu}_{dt}^{syn}$, under Model 0 (left) and of the EBP $\hat{\mu}_{dt}$ under Model ST1 (right).
A clear improvement is achieved when one uses a more complex model, since the Pearson residuals are closer to 0.
In this case, the EBP of $\mu_{dt}$ under Model ST1 is more competitive than the plug-in predictor under Model $0$, since the empirical MSEs of Pearson residuals are $0.454$ and $1.247$ respectively.

\begin{figure}[h]
  \centering
  \begin{tabular}{ccc}
  \hspace*{-1cm}
  \includegraphics[width=6cm, height=5.6cm]{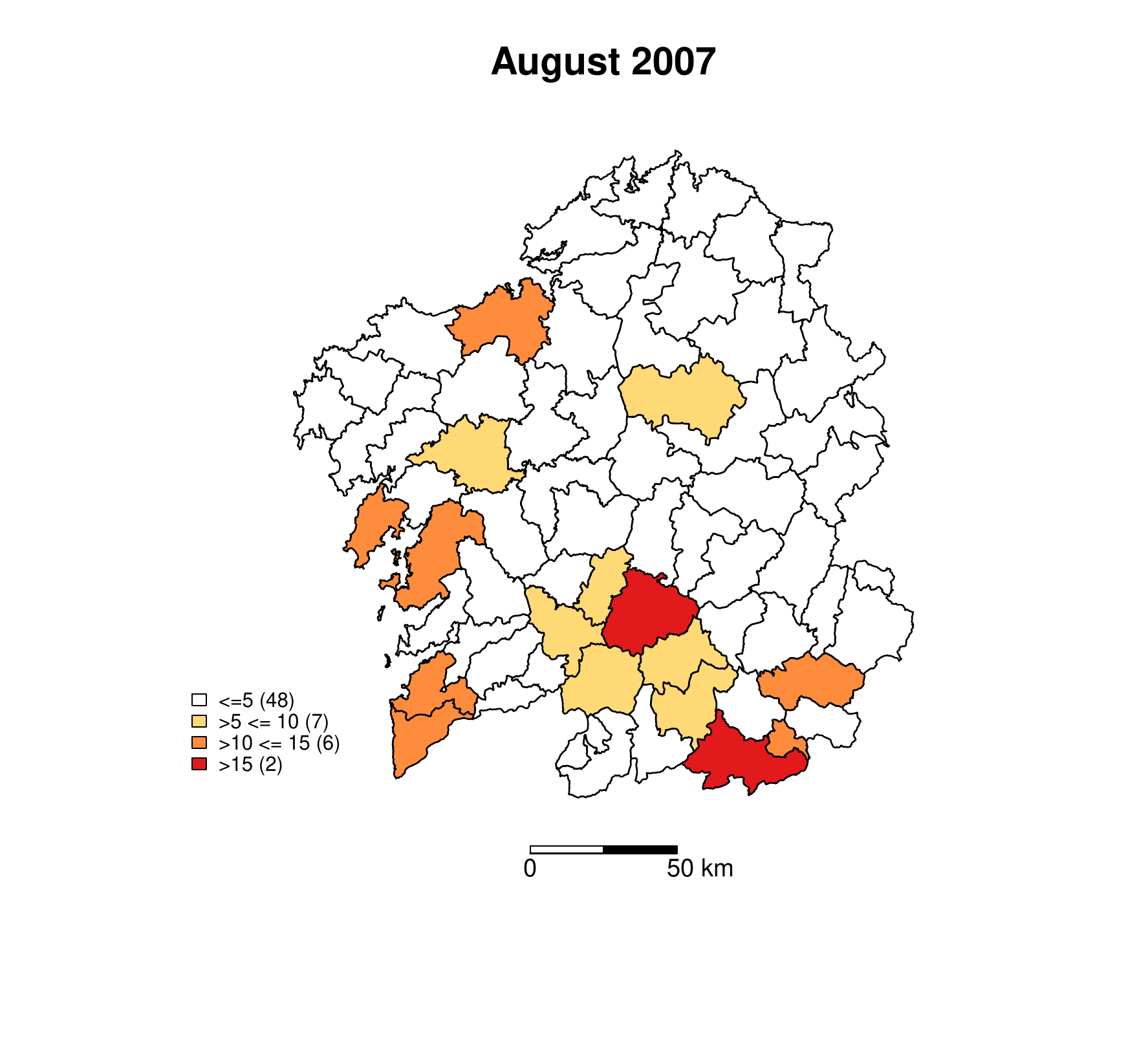}
  \hspace*{-1cm}
  \includegraphics[width=6cm, height=5.6cm]{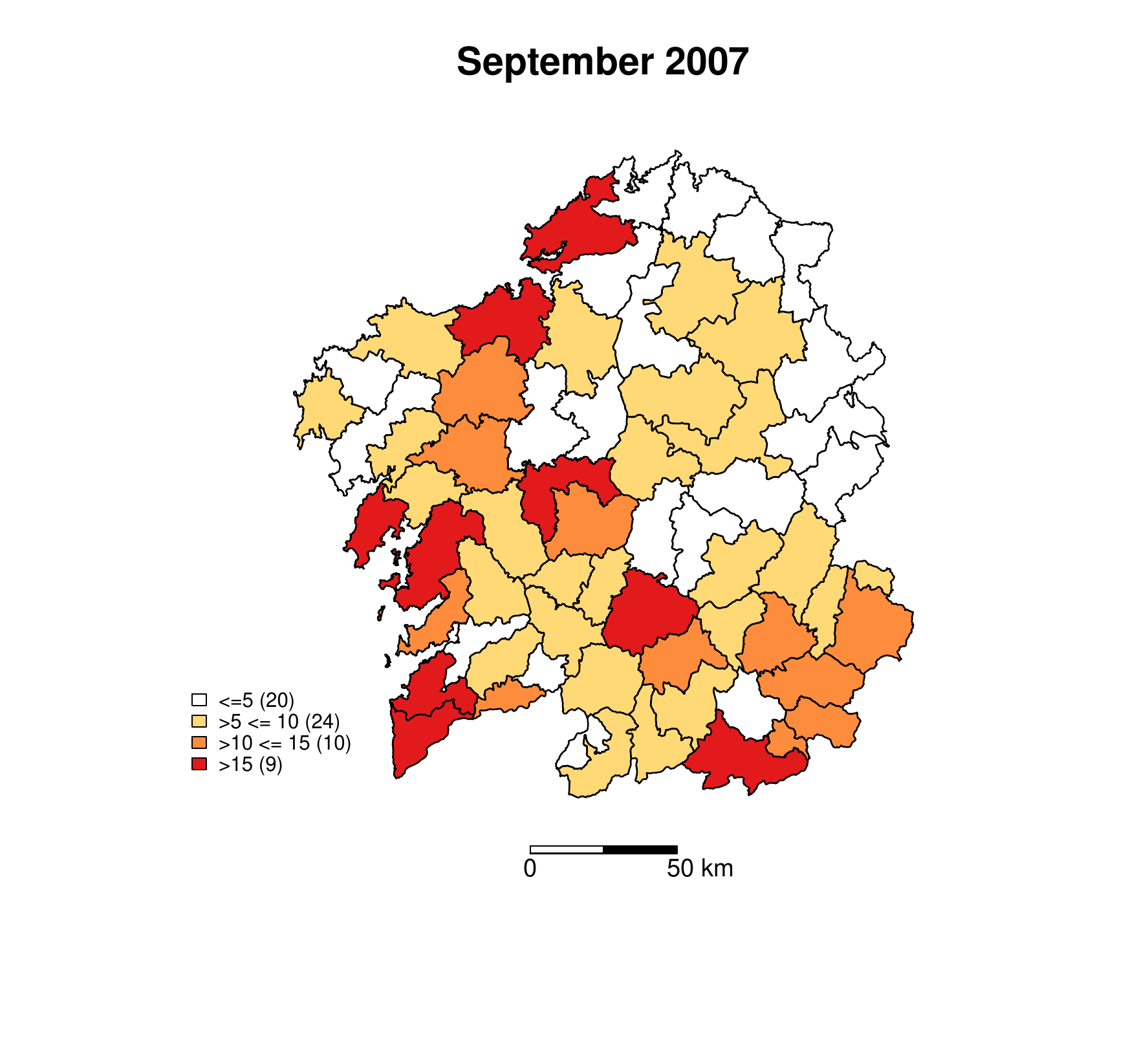}
  \hspace*{-1cm}
  \includegraphics[width=6cm, height=5.6cm]{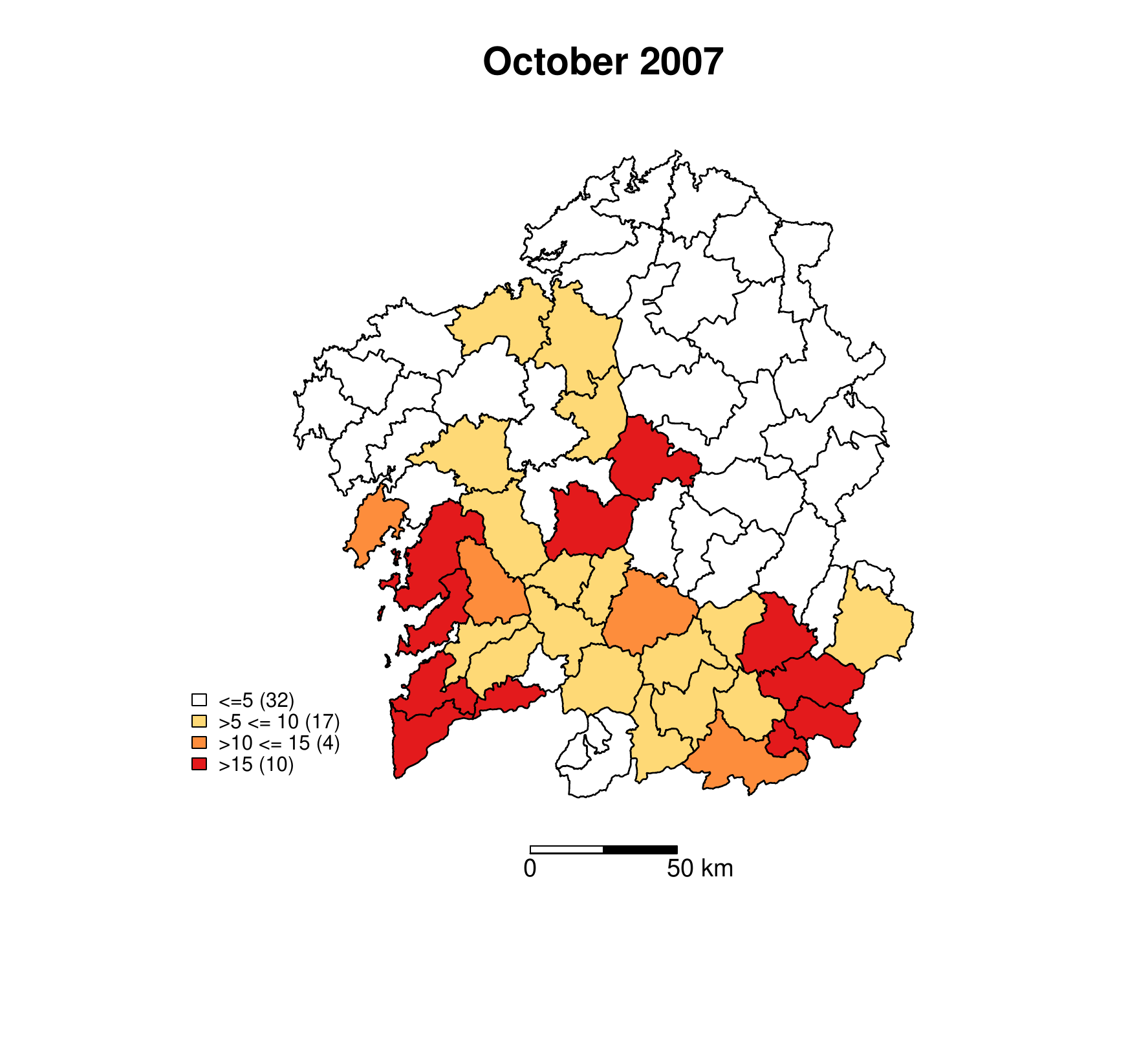} \\[-.8cm]
  \hspace*{-1cm}
  \includegraphics[width=6cm, height=5.6cm]{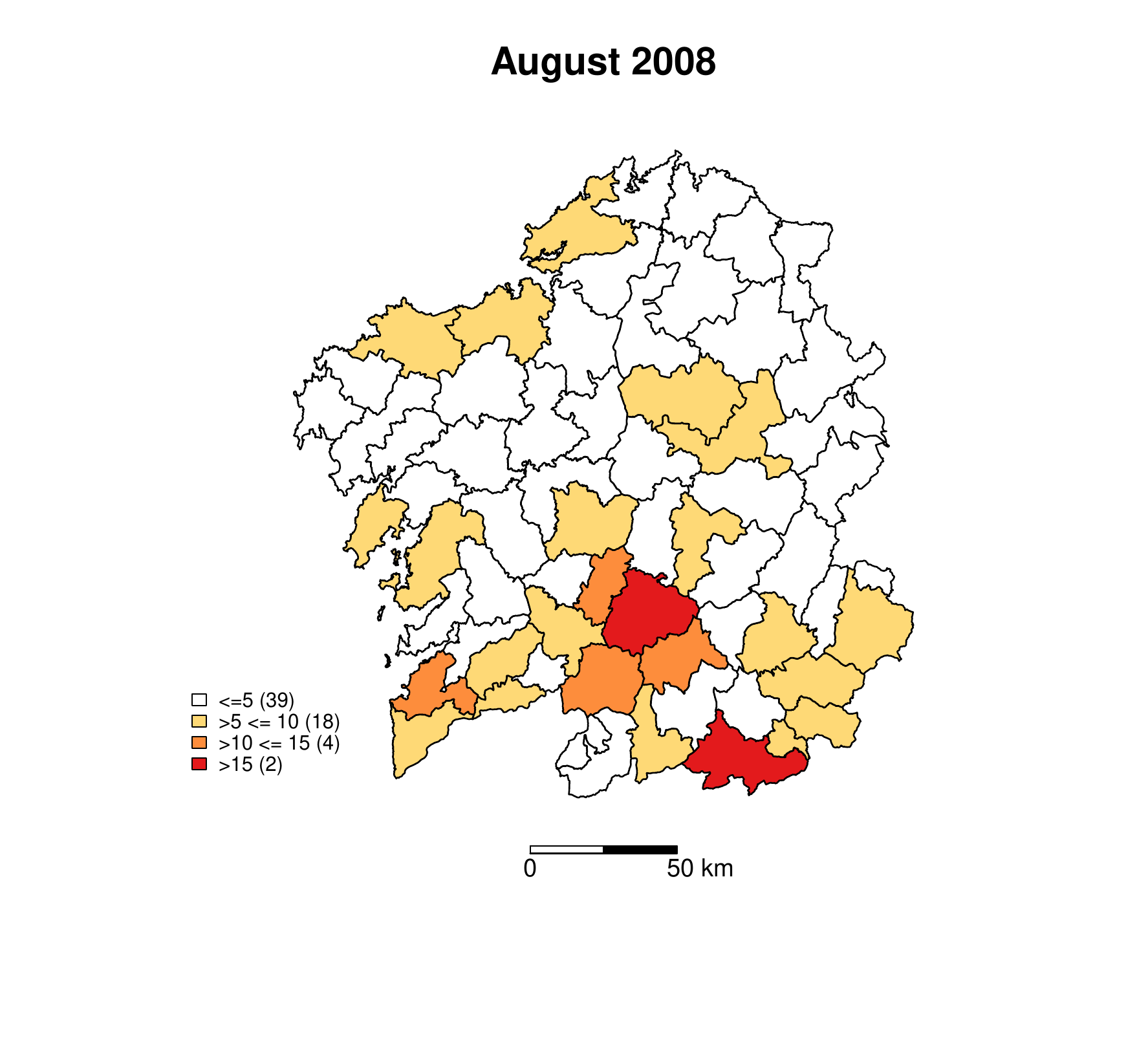}
  \hspace*{-1cm}
  \includegraphics[width=6cm, height=5.6cm]{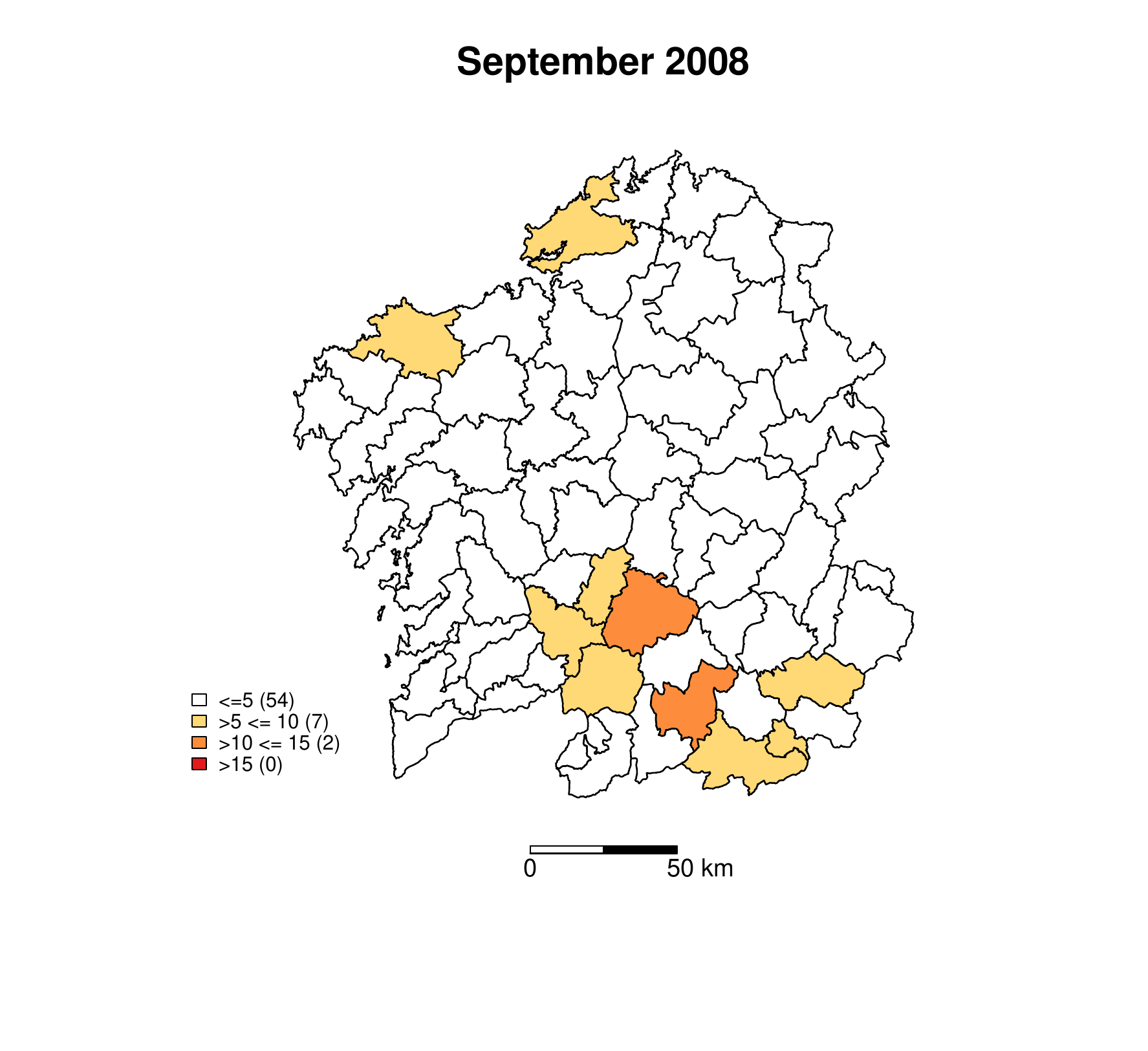}
  \hspace*{-1cm}
  \includegraphics[width=6cm, height=5.6cm]{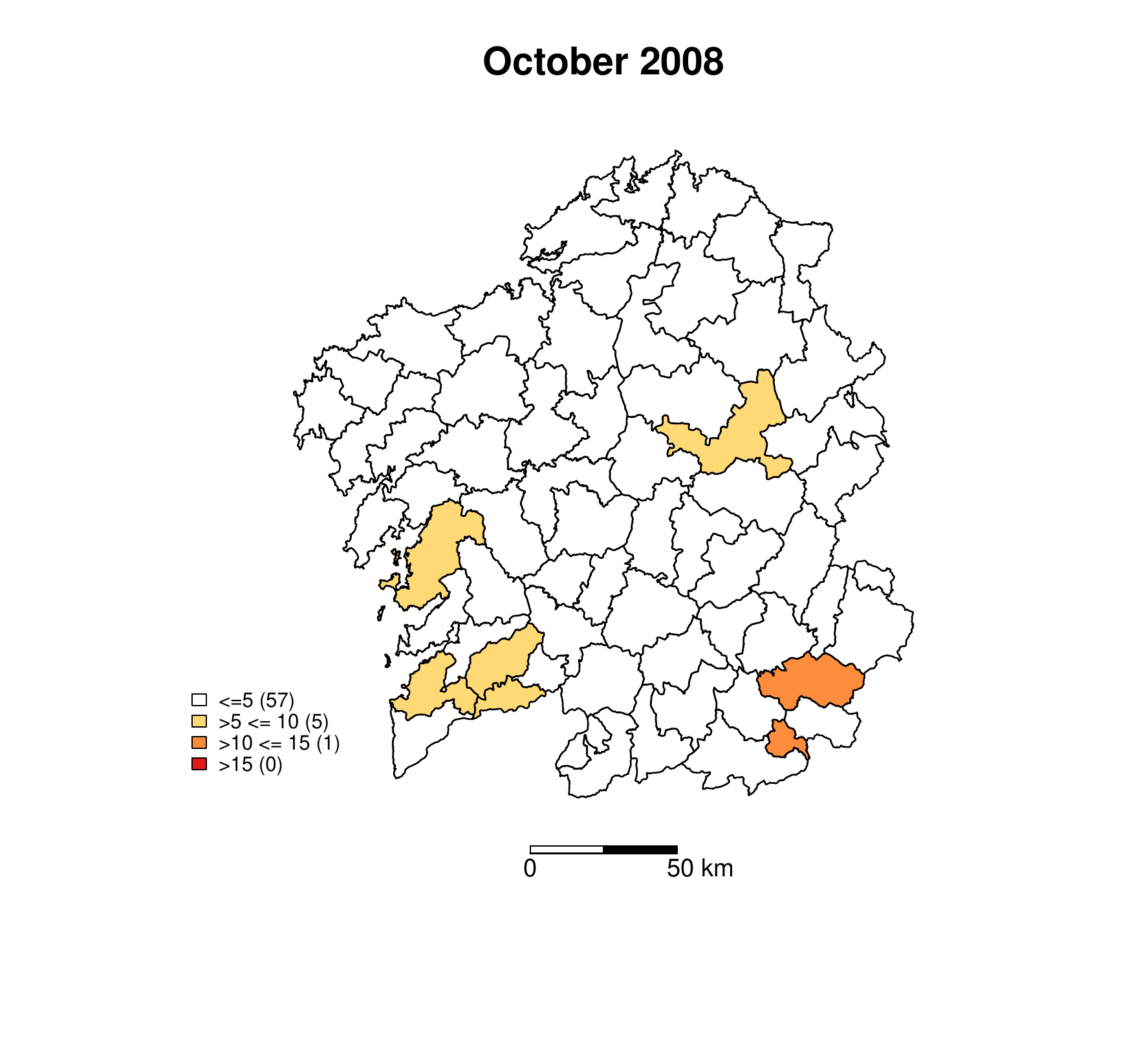}
  \end{tabular}
  \vspace*{-.5cm}
  \caption{Estimated fires from August to October in 2007-2008.}
  \label{sec6-yPlugAPoissTAR1Map}
\end{figure}

Figure \ref{sec6-yPlugAPoissTAR1Map} maps the EBPs of numbers of wildfires by forest areas under the fitted Model ST1.
We group the forest areas by number of fires with cuts in 5, 10 and 15.
The legends present, between parentheses, the number of forest areas in each group.
We take the same months between August and October.
The figure suggests that the highest number of wildfires are concentrated in western coastal areas and in the south of the region.
On the other hand, regarding the temporal behaviour, the highest number of fires is found in the months of $2007$ (specially in September and October, being a year with the summer delayed),
while in $2008$ there was an impressive decrease, because it was a particularly rainy summer (information contrasted with MeteoGalicia).

\begin{figure}[H]
  \centering
  \begin{tabular}{c}
  \includegraphics[width=11cm, height=6cm]{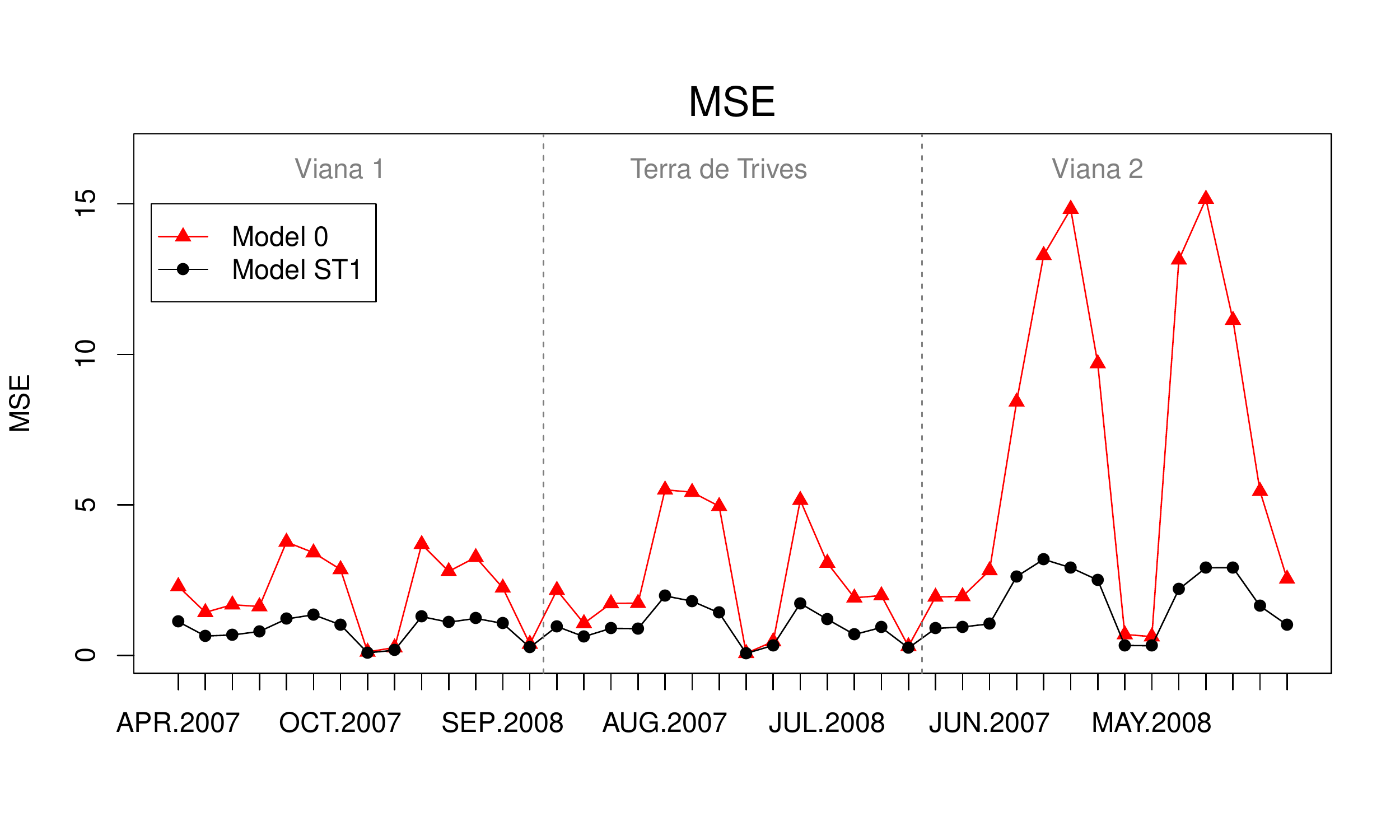}
  \end{tabular}
   \vspace*{-.5cm}
  \caption{Bootstrap MSE estimates for the three areas with highest number of fires.}
  \label{sec6-MSE}
\end{figure}

Figure \ref{sec6-MSE} plots the bootstrap MSEs, described in Section \ref{sec3.3}, for the three areas with highest number of fires (Viana 1, Terra de Tribes and Viana 2).
The number of bootstrap replicates is $B=500$. The mean of the MSEs for the three areas is $3.978$ in Model 0 and $1.219$ in Model ST1.
Then, a clear improvement is achieved when one uses the area-level spatio-temporal Poisson mixed model ST1.
%
%
\section{Application to poverty data}\label{sec.povdata}
%
%
Based on Model S1, as a particular case of Model ST1, this section applies the developed statistical methodology to estimating women poverty proportions by counties of Galicia during 2013.
The data are taken from the Spanish living conditions survey (SLCS), which provides annual information about the household income received during the year prior to that of the interview.
Since there is only one time period (year 2013), this section removes the $t$ index.

For every individual, the equivalent personal income is obtained by dividing the annual household net income by the equivalent total of household members, which is obtained as a weighted sum. The poverty line is defined as a percentage (currently Eurostat fixed it in 60\%) of the median of the equivalent personal incomes in the whole country. A person is defined as poor (poverty=1) if his/her equivalent personal income is lower than the poverty line and is defined as not poor (poverty=0) otherwise. The poverty proportion of a territory is the average of the poverty variable.
If the average is restricted to women, then the corresponding poverty proportion for women is obtained.

Spain is hierarchically divided in autonomous communities, provinces, counties and municipalities.
The SLCS is designed to obtain reliable statistics for autonomous communities.
Therefore, the usual direct estimators are not precise enough for studying the poverty at a more disaggregated level, like provinces or counties (small areas).
We use the introduced SAE methodology for constructing model-based predictors that give more accurate estimations of women poverty proportions for each county area and find differences between them.

\begin{figure}[h]
  \centering
  \begin{tabular}{c}
  \includegraphics[width=7cm, height=7cm]{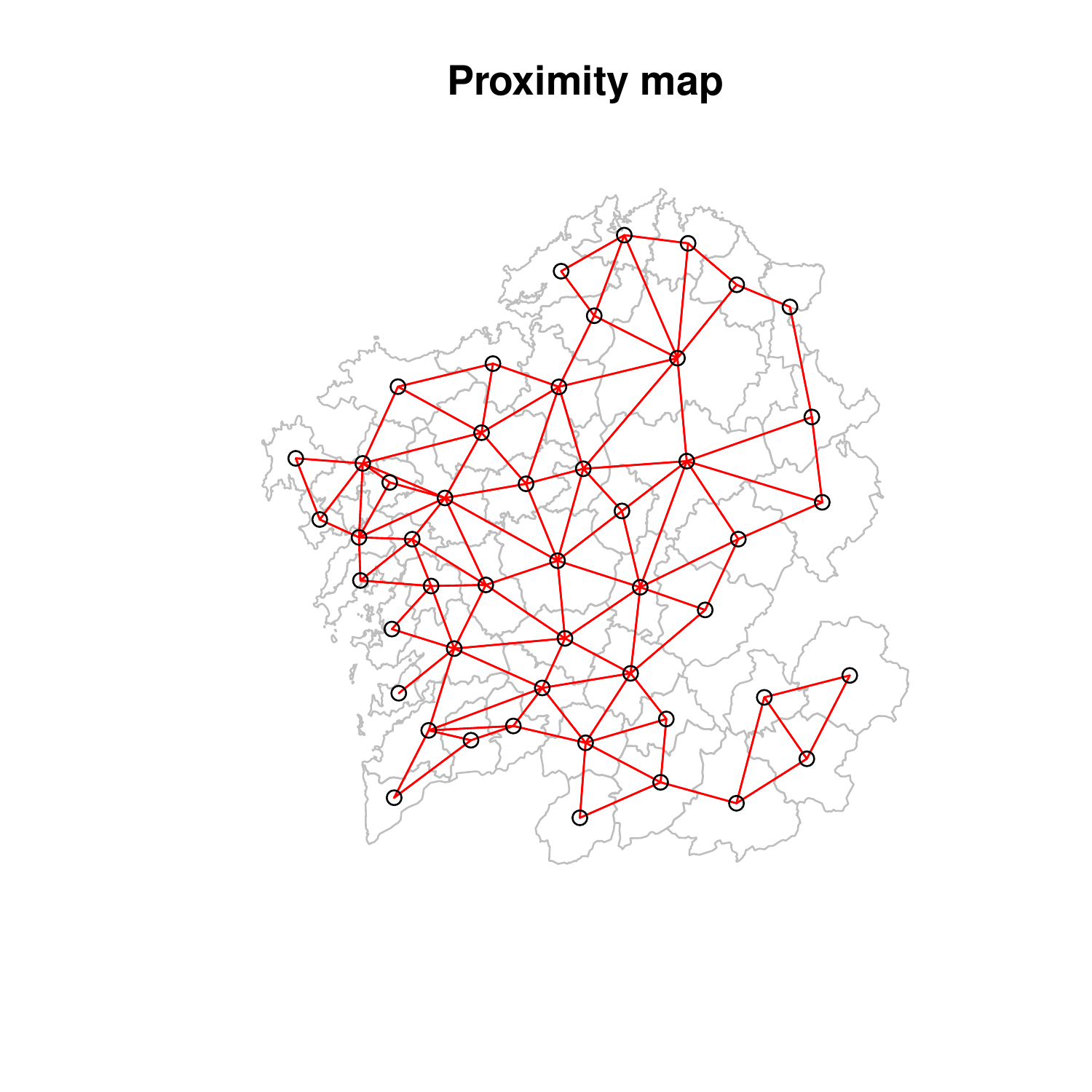}
  \end{tabular}
  \vspace*{-1.5cm}
  \caption{Proximity map for each domain $d$ ($d=1,\ldots, D$).}
  \label{LCSfig1}
\end{figure}

In Galicia there are 53 counties, but in four of them there are no available data. Therefore, the number of considered domains is $D=49$. The performance of Model S1 depends on the choice of the proximity matrix $\WW$. Three different choices are tested: common borders,  distances and $k$-nearest neighbours. In the first option (common border), two domains are neighbours if they have a common delimitation. The last two options consider the Euclidean distance between the centroids of the counties. The second option sets up a proximity measure by taking the inverse of the distance between domains. The last option applies $k$-nearest neighbours with $k=2$ and $3$.
After analysing the different possibilities in terms of model diagnostics, the first option is selected because it is the one giving the best results.
Figure \ref{LCSfig1} shows the proximity map that determines the proximity matrix $\WW^0$, i.e. it provides for each domain, which are its neighbours.
See Section \ref{sec2} for more details on the construction of the proximity matrix $\WW^0$ and $\WW$.

The target variable, $y_d$, counts the number of women under the poverty line in the domain $d$ and $\nu_d$ is the corresponding sample size of women.
The minimum, median and maximum values of $\nu_d$ are $19$, $152$ and $1384$, respectively. The minimum has been reached in the south east of the region,
while the median and maximum belong to the south west.
The auxiliary variables are the proportions of women who are unemployed ($lab2$) and who have completed university education ($edu3$).
These variables are calculated by averaging direct estimates of the four Spanish employment surveys carried out in 2013, all of them having bigger sample sizes than the SLCS.
We first fit Model 1 to the data $(y_d,\nu_d,\xx_d)$, $d=1,\ldots,D$ and we apply Moran's I test for spatial autocorrelation.
As the obtained $p$-value is lower than $0.001$, we assume that $y_d$, $d=1,\ldots, D$, follows Model S1 and we fit this model to the data.

Table \ref{Descriptab} provides a descriptive analysis for the logarithm of the response variable, $\log y$, and the considered covariates.
Specifically, it presents their mean, standard deviation (sd), median, minimum and maximum values (min and max) and the correlation (cor) between the covariates and $\log y$.
\begin{table}[h]
\centering
\caption{Descriptive analysis.}
\label{Descriptab}
\begin{tabular}{lrrrrrr}
\toprule
  Variable & mean & sd & median & min & max & cor \\
\toprule
  $\log y$ & 3.13 & 1.15 & 3.18 & 0.00 & 5.51 & \multicolumn{1}{c}{$-$} \\
  $lab2$   & 0.10 & 0.04 & 0.10 & 0.02 & 0.21 & 0.32 \\
  $edu3$   & 0.15 & 0.07 & 0.14 & 0.03 & 0.33 & 0.42 \\
\bottomrule
\end{tabular}
\end{table}

Table \ref{LCStab1} presents the significant estimates ($p$-value $<0.05$) of the fixed regression parameters under Model S1 and their standard errors, $z$-values and $p$-values.
The spatial autocorrelation parameter is estimated by applying Moran's I statistic (\ref{MoranI}) over the Pearson residuals of Model 0 and the remaining model parameters are given as a
solution of the system formed by the first $p+1$ MM equations in (\ref{secMM-MM}).

\begin{table}[h]
\centering
\caption{MM estimates of regression parameters under Model S1.}
\label{LCStab1}
\begin{tabular}{lrrrr}
\toprule
  Variable & Estimate & s.e. & $z$-value & $p$-value \\
\midrule
  $Intercept$ & -1.8803 & 0.1515 &-12.4086 & $<$ 0.001 \\
  $lab2$      &  2.9848 & 1.2097 &  2.4689 & 0.0136 \\
  $edu3$      & -1.3809 & 0.5033 & -2.7445 & 0.0061 \\
\bottomrule
\end{tabular}
\end{table}

Taking into account the signs of the estimates, the auxiliary variable $lab2$ (proportion of unemployed women), is directly related to the response variable while $edu3$ (proportion of women with university level of education), helps to decrease the counts of women under the poverty line.
Each domain $d$, $d=1,\ldots,D$, has a random intercept with distribution $N(0, \phi^2)$, where $\hat{\phi}=0.130$. The $95\%$ percentile bootstrap confidence interval for the standard deviation parameter is $(0.001, 0.331)$. The estimated autocorrelation parameter is $\hat{\rho} = 0.324$.
To test the null hypothesis $H_0: \phi^2=0$,  Algorithm 1 of Section \ref{sec2} is applied. The obtained $p$-value is $0.018$.
Then, the null hypothesis is rejected at the level $\alpha=0.05$.
The Algorithm 2 of Section \ref{sec2} adapted to case $T = 1$, and which compares Model S1 vs. Model 1, is applied to test $H_0: \rho=0$.
The obtained bootstrap $p$-value is $0.001$. Taking $\alpha=0.05$, the bootstrap test concludes that the autocorrelation parameter $\rho$ is significantly different from 0.
Therefore, it recommends fitting Model S1 to the data, instead of Model 1.

\begin{figure}[h]
  \centering
  \begin{tabular}{cc}
  \includegraphics[width=6.5cm, height=6cm]{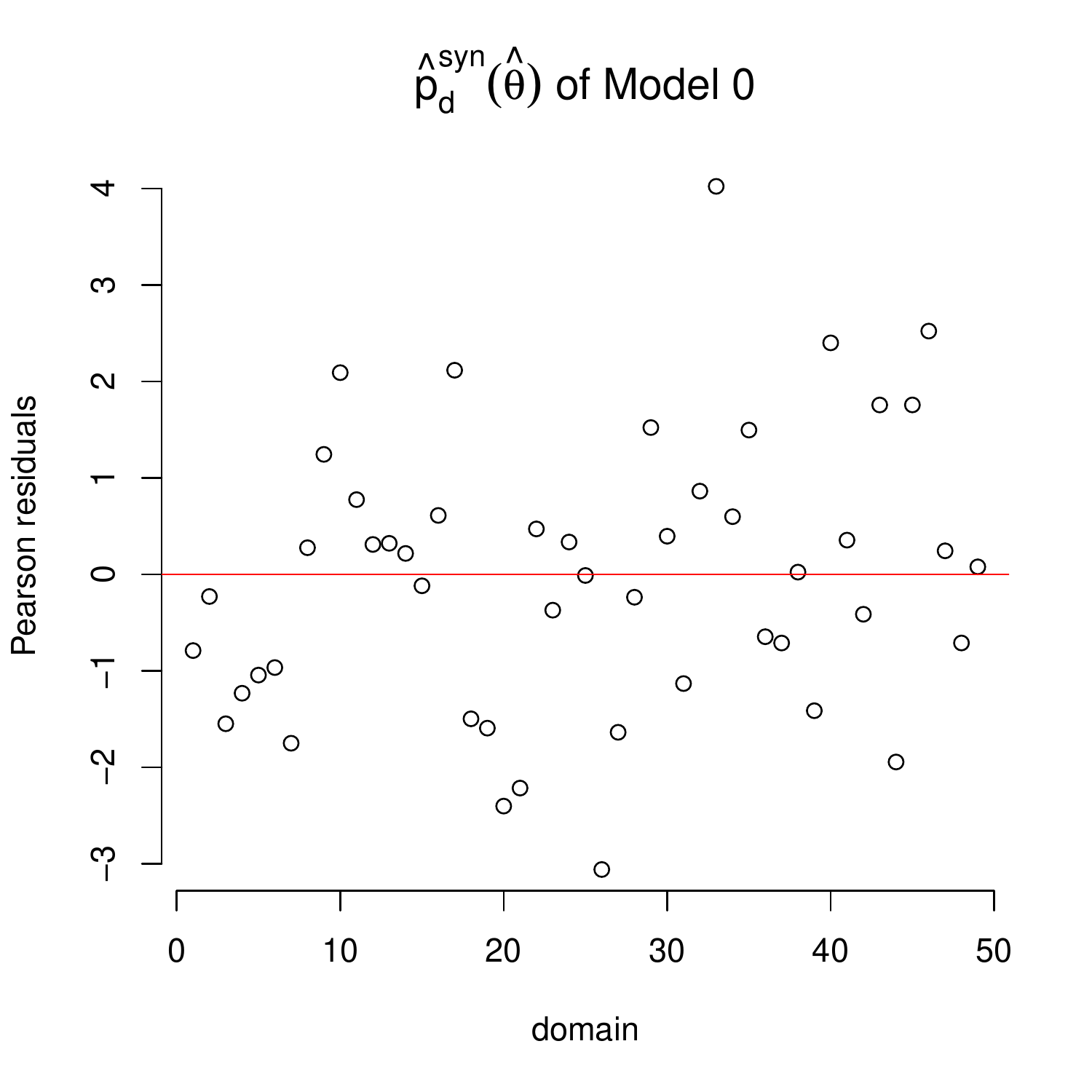}&
  \includegraphics[width=6.5cm, height=6cm]{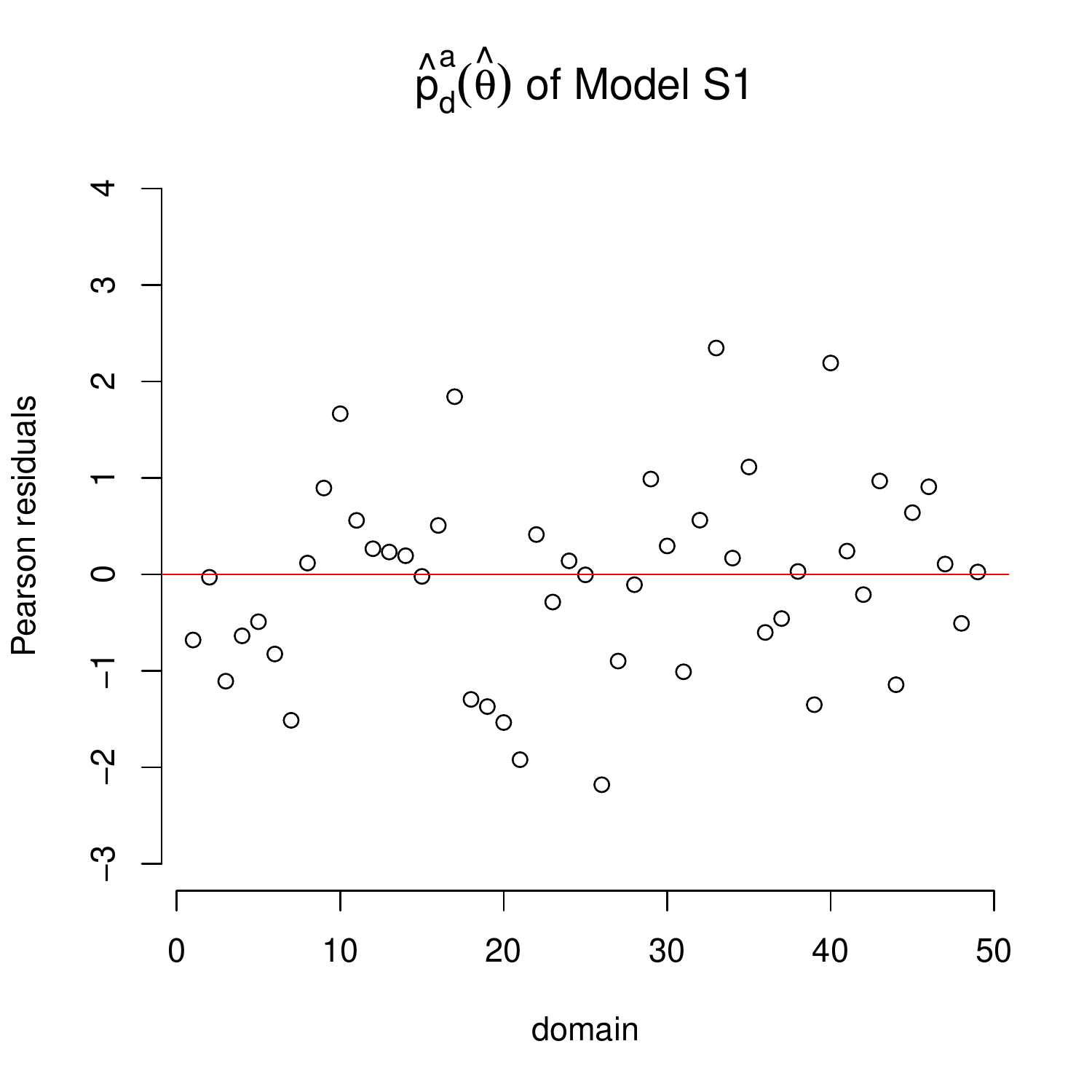}
  \end{tabular}
  \vspace*{-.5cm}
  \caption{Pearson residuals of the synthetic estimator based on Model 0 (left) and of the EBP approximation based on Model S1 (right).}
  \label{LCSfig2}
\end{figure}

We compare the performance of Model S1 and Model 0, with only fixed-effects (or equivalently, with $\phi=0$).
For this sake, we fit Model 0 to the same data as Model S1. Figure \ref{LCSfig2} plots the Pearson residuals
$$
r_d^P=\frac{y_d-\nu_d\hat{p}_d}{\sqrt{\nu_d\hat{p}_d}},\quad d=1,\ldots,D,
$$
of the synthetic estimator $\hat{p}_d^{syn}(\hat{\bbeta})=\exp\{\xx_d\hat{\bbeta}\}$ based on Model 0 (left), and of the EBP approximation $\hat{p}_d^{a}(\hat{\ttheta})$ based on Model S1 (right).
In both cases, the distribution of the Pearson residuals is symmetrical around 0. In addition, the plots suggest a clear improvement when one uses an area-level Poisson mixed model that incorporates SAR(1) domain effects, since it is able to better capture the underlying spatial correlation structure.
Figure \ref{LCSfig2} shows that most of the residuals of Model S1 take values in the interval $(-2,2)$, while the residuals of Model 0 take values mainly in the interval $(-3,3)$.
In addition the tested hypotheses on $\rho$ and $\phi$, the conclusion is again that Model S1 is more appropriated to fit the women poverty data in Galicia by counties in $2013$.

\begin{figure}[h]
  \centering
  \begin{tabular}{cc}
  \includegraphics[width=6.5cm, height=6cm]{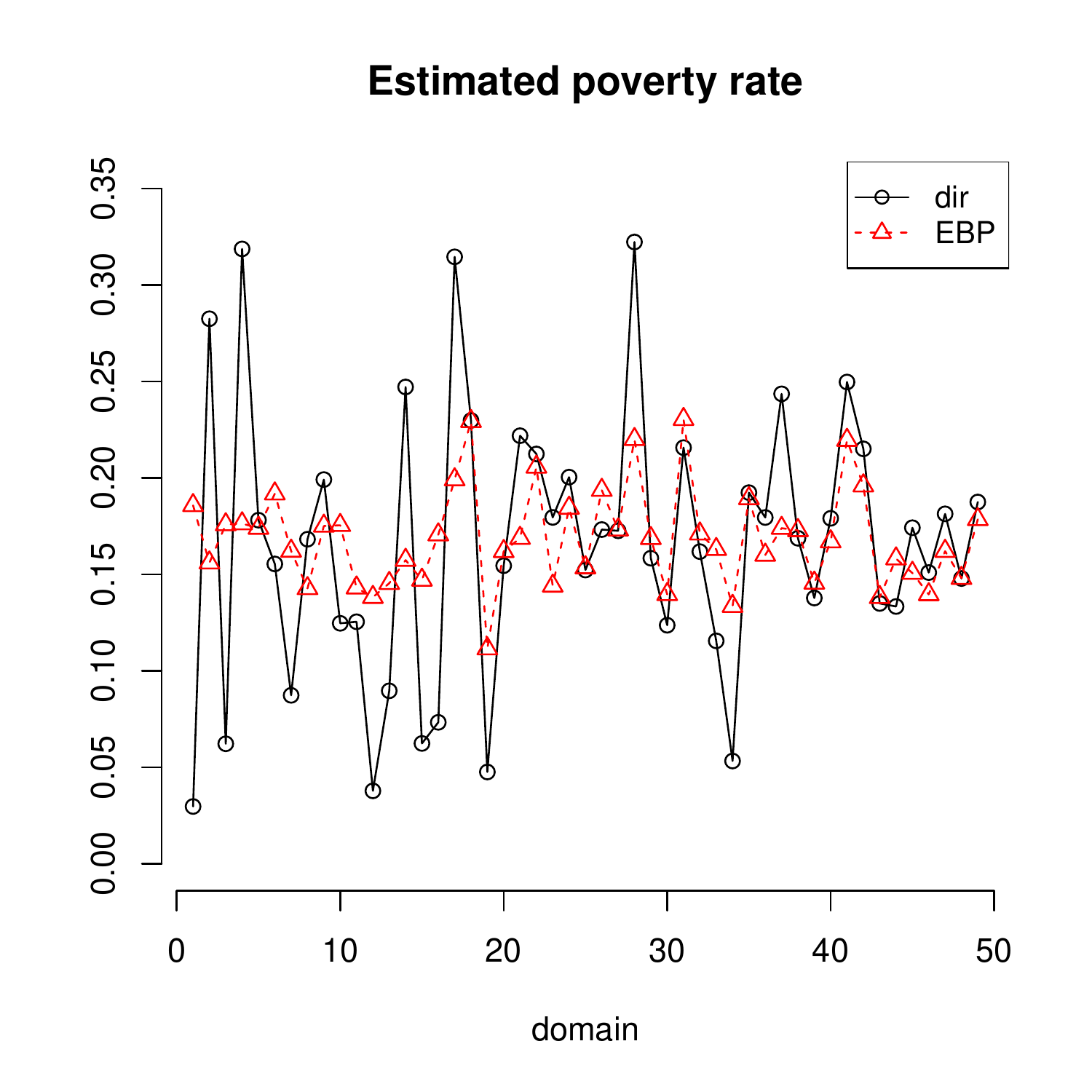}&
  \includegraphics[width=6.5cm, height=6cm]{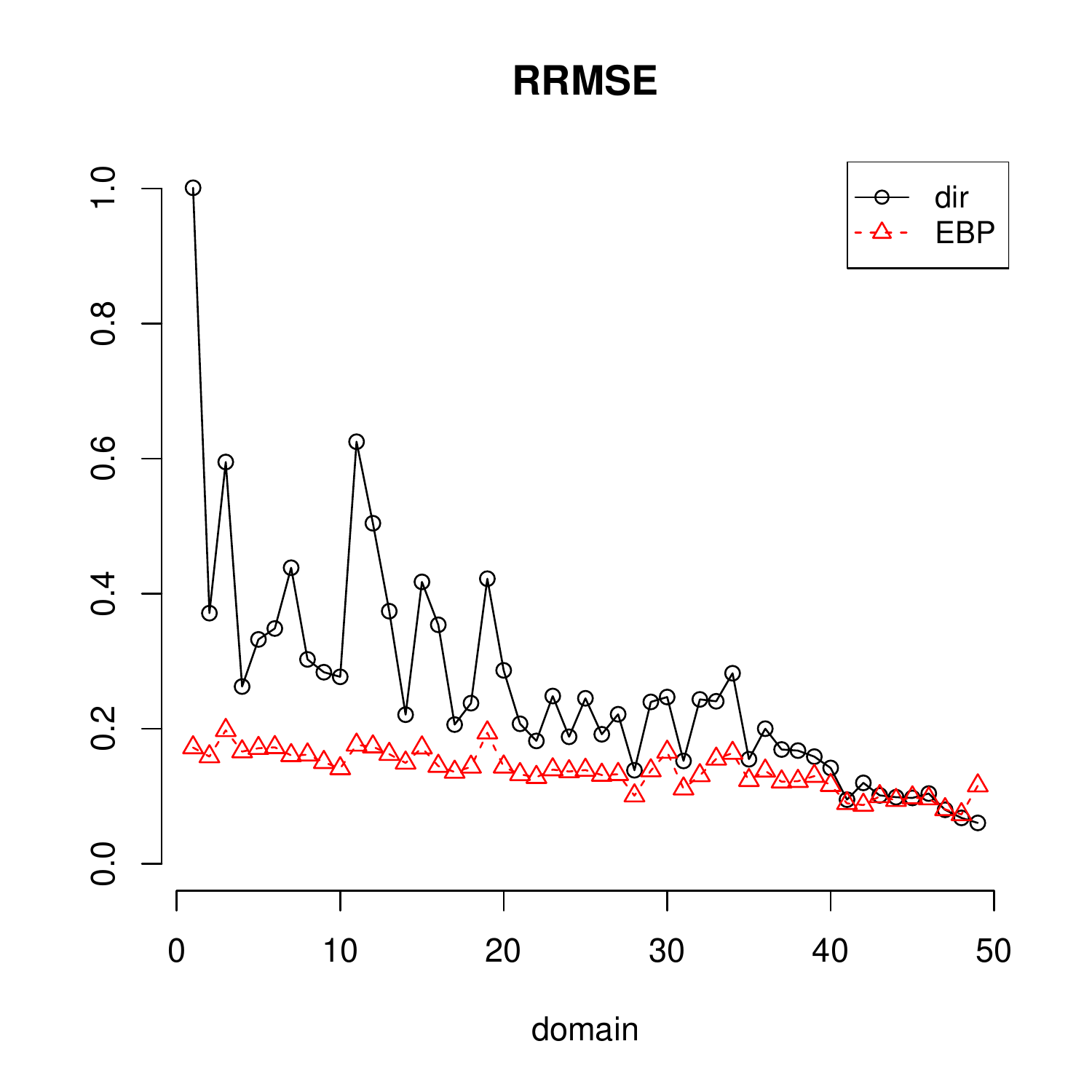}
  \end{tabular}
  \vspace*{-.5cm}
  \caption{Direct estimates and EBPs of poverty proportions $p_d$ (left) and relative root-MSEs (right) for women in $2013$.}
  \label{LCSfig3}
\end{figure}

Figure \ref{LCSfig3} (left) compares the behaviour of the EBP and direct estimations.
The direct estimators are calculated by using the H\'ajeck formula with the officially calibrated sampling weights.
The domains are sorted by the sample sizes $\nu_d$'s.
The direct estimators present oscillations of large amplitude, while the EBPs have a smoother behaviour, which is something preferred by the statistical offices when publishing estimations. As the sample size increases, both sets of estimates tend to overlap.

Figure \ref{LCSfig3} (right) plots the relative root-MSEs (RRMSE) of the EBPs based on Model S1 and the relative root-variances of the direct estimators.
The RRMSEs of the EBPs are estimated by using the bootstrap procedure of Section \ref{sec3.3} with $B=500$ replicates.
For the direct estimators, we apply the formula (10.3.8) of S\"arndal et al. (1992).
The direct estimates have high variability, specially for small sample sizes. As above, when the sample size $\nu_d$ increases, both accuracy measures follow the same pattern.
The averages of the relative root-variances of the direct estimator and of the RRMSEs of the EBP are $0.2595$ and $0.1323$, respectively.
According to these results, we conclude that the EBP performs better.

\begin{figure}[h!]
  \centering
  \begin{tabular}{cc}
  \includegraphics[width=9cm, height=8cm]{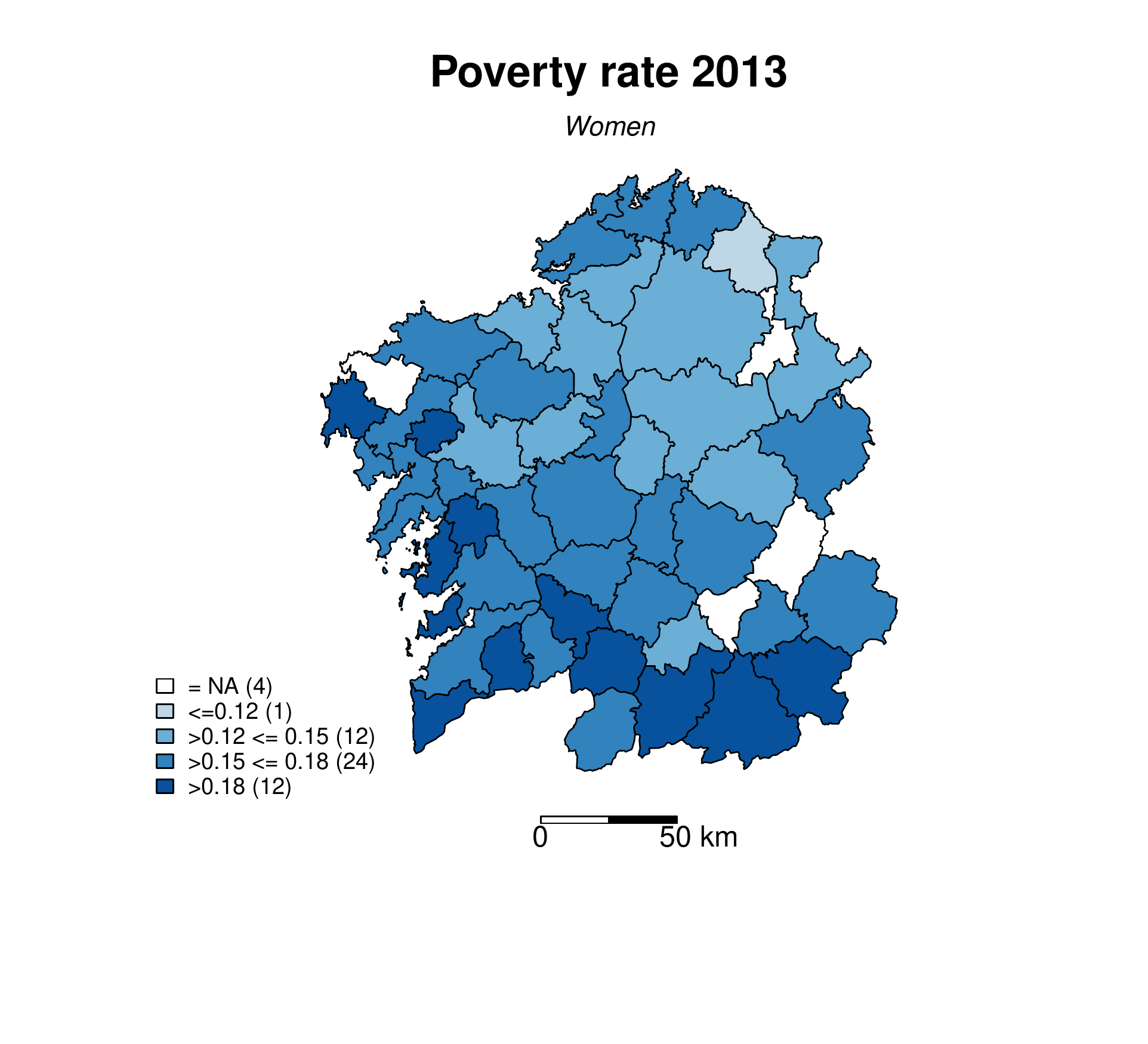}&
  \hspace*{-1cm}
  \includegraphics[width=9cm, height=8cm]{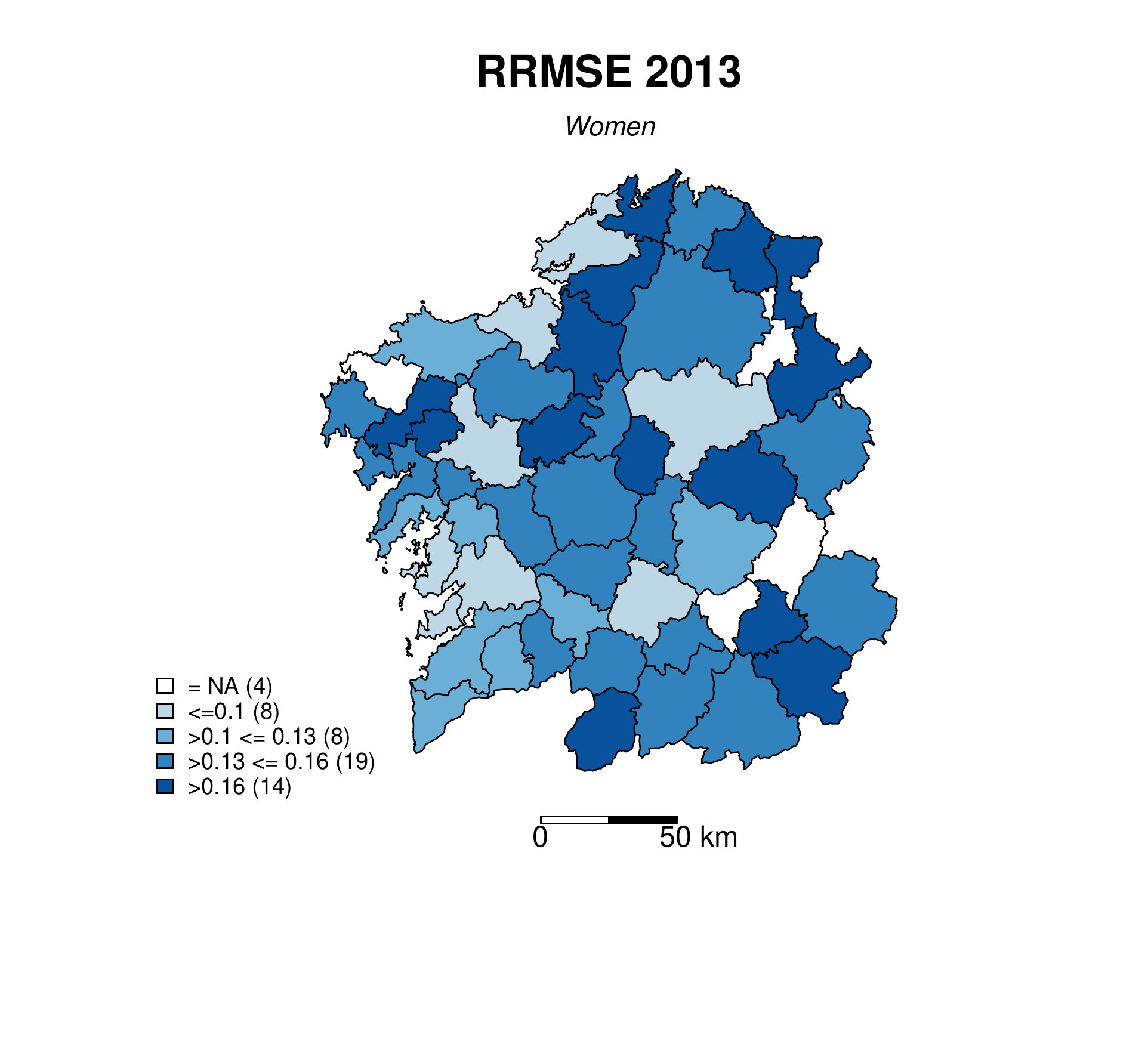}
  \end{tabular}
  \vspace{-1.2cm}
  \caption{Poverty proportion EBPs for women based on Model S1 (left) and RRMSEs (right) in Galicia during $2013$.}
  \label{LCSfig4}
\end{figure}

Figure \ref{LCSfig4} (left) maps the EBPs of $p_d$ for women in $2013$. The regions where there is no data, are in white.
Model S1 gives the following predictions of women poverty proportions: 1 county with poverty proportion $p_d \leq 0.12$, 12 counties with $0.12 < p_d \leq 0.15$, 24 counties with $0.15 < p_d \leq 0.18$ and 12 counties with $p_d > 0.18$. Highest levels of poverty are found in the south and west of the community. On the other hand, the counties with the lowest estimated poverty proportion are located in the north-east of the region.

Figure \ref{LCSfig4} (right) maps the RRMSE estimates of the EBP of $p_d$ by counties in $2013$. There are 8 counties with RRMSE $\leq 10\%$, 8 counties with $10\% <$ RRMSE $\leq 13\%$, 19 counties with $13\% <$ RRMSE $\leq 16\%$ and 14 counties with RRMSE $> 16\%$. The highest values are found in the north-east of the region. Their minimum and maximum are $8.82\%$ and $18.49\%$, respectively. As the highest RRMSE is lower than $20\%$, these estimates could be accepted for publication by statistical offices.

%
%
%
%
\section{Concluding remarks}\label{sec.conclus}
%
This paper introduces Model ST1, which is an area-level spatio-temporal Poisson mixed model with SAR(1)-correlated domain random effects and independent domain-time random effects.
It contains the area-level Poisson mixed model with SAR(1) domain effects (Model S1) and the area-level Poisson mixed model with independent time effects (Model T1) as particular or limit cases.
The MM algorithm is obtained to fit the model parameters. It is based on the method of simulated moments proposed proposed by Jiang (1998).

Two predictors of the target parameter $p_{dt}$ are proposed.
They are the EBP and the plug-in predictor.
The EBP is asymptotically unbiased with minimum MSE and it is computationally faster than the plug-in predictor, since the last one requires calculating two EBPs ($\hat{\vv}_1$ and $\hat{\vv}_2$).
Further, an approximation to the EBP is proposed to avoid the computational burden.

The developed methodology is applied to the data set of wildfires in Galicia by forest areas and months during 2007-2008.
Because of their better properties, the EBP is employed.
This predictor is also compared against the synthetic estimator obtained under Model 0, with only fixed effects.
A clear improvement is achieved when one uses a more complex model incorporating random effects.
In addition, this paper recommends using Model ST1 to analyse wildfires in Galicia since the bootstrap test of Algorithm 2 yielded $p$-values close to 0.


The conclusion is  that forest fires tend to be concentrated in coastal areas and in the south of the region.
An important increase in wildfires is observed in September and October $2007$.
The introduced bootstrap MSE estimator is considered as accuracy measure of the proposed EBP.
It is achieved a clear improvement when using the proposed model and estimator against the classical Poisson regression model.

We also use the EBP approximation for estimating women poverty proportions in Galician counties.
The data are taken from the 2013 SLCS.
As both spatial correlation tests, the  Moran's I test and the modified Algorithm 2, indicates spatial correlation, we fit Model S1 to the data.
In addition, the proposed model-based predictor is compared against the direct estimator.
The EBPs of the women poverty proportion are smoother.
As the RRMSE of the direct estimator is too high when the sample size $\nu_d$ is small, it is preferable to use the EBP.
The estimates based on  Model S1 suggests that the highest levels of women poverty are found in the south and west of the region.
The average percentage of women poverty is $16.89\%$ and its average error is $13.23\%$.

It is worth mentioning the programming problems of the predictors constructed under the introduced spatio-temporal models, which leads us to use numerical approaches that introduce an additional source of error.
However, if there is a spatial correlation, the approximate EBPs behave better than other predictors that do not take into account this information.

%
%
%
\appendix
\section{Appendix: MM equations}\label{apMM}
First, we recall that the moment generation function of $Y\sim N(\mu,\sigma^2)$ is
\begin{equation*}\label{chap3:secMM-MGFNormal}
\Psi(t;\mu,\sigma^2)= \mathbb{E}\left[e^{tY}\right]=\exp\big\{\mu t+\frac12\,\sigma^2 t^2\big\}.
\end{equation*}

For ease of exposition, the elements of $\GGamma$ and its derivatives are denoted by $\gamma_{d_1d_2}=\gamma_{d_1d_2}(\rho)$ and $\dot{\gamma}_{d_1d_2}=\dot{\gamma}_{d_1d_2}(\rho)$, respectively.
The calculations start with the first $p$ MM equations. The expectation of $y_{dt}$ is
\begin{align*}
\mathbb{E}_{\ttheta}[y_{dt}]&=\mathbb{E}_v\big[E_{\ttheta}[y_{dt}|\vv]\big]=\mathbb{E}_v[\nu_{dt}p_{dt}]=
\mathbb{E}_v\left[\nu_{dt}\exp\left\{\xx_{dt}\bbeta+\phi_1 v_{1,d}+\phi_2 v_{2,dt}\right\}\right]
\\[.15cm]
&=\int_{-\infty}^{\infty}\int_{-\infty}^{\infty}\nu_{dt}\exp\left\{\xx_{dt}\bbeta+\phi_1 v_{1,d}+\phi_2 v_{2,dt}\right\}\,
f(v_{1,d})f(v_{2,dt})\,dv_{1,d}dv_{2,dt}
\\[.15cm]
&=
\int_{-\infty}^{\infty}\nu_{dt}\exp\Big\{\xx_{dt}\bbeta+\frac12\phi_2^2+\phi_1 v_{1,d}\Big\}f_v(v_{1,d})\,dv_{1,d}
\\[.15cm]
&=
\nu_{dt}\exp\Big\{\xx_{dt}\bbeta+\frac12\big(\phi_1^2\gamma_{dd}+\phi_2^2\big)\Big\}.
\end{align*}
Therefore, the first $p$ MM equations are
\begin{align*}
f_k(\ttheta)&=\frac{1}{DT}\sum_{d=1}^D\sum_{t=1}^{T}\nu_{dt}\exp\Big\{\xx_{dt}\bbeta+\frac12\big(\phi_1^2\gamma_{dd}+\phi_2^2\big)\Big\}x_{dtk}
\\[.15cm]
&-\frac{1}{DT}\sum_{d=1}^D\sum_{t=1}^{T}y_{dt}x_{dtk},\,\, k=1,\ldots, p.
\end{align*}
The derivatives of $\mathbb{E}_{\ttheta}[y_{dt}]$ are
\begin{align*}
\frac{\partial \mathbb{E}_{\ttheta}[y_{dt}]}{\partial\beta_k}&=\nu_{dt}\exp\big\{\xx_{dt}\bbeta+\frac12(\phi_1^2\gamma_{dd}+\phi_2^2)\big\} x_{dtk},
\\[.15cm]
\frac{\partial \mathbb{E}_{\ttheta}[y_{dt}]}{\partial\phi_1}&=\nu_{dt}\exp\big\{\xx_{dt}\bbeta+\frac12(\phi_1^2\gamma_{dd}+\phi_2^2)\big\} \phi_1\gamma_{dd},
\\[.15cm]
\frac{\partial \mathbb{E}_{\ttheta}[y_{dt}]}{\partial\phi_2}&=\nu_{dt}\exp\big\{\xx_{dt}\bbeta+\frac12(\phi_1^2\gamma_{dd}+\phi_2^2)\big\} \phi_2,
\\[.15cm]
\frac{\partial \mathbb{E}_{\ttheta}[y_{dt}]}{\partial\rho}&=\frac12\nu_{dt}\exp\big\{\xx_{dt}\bbeta+\frac12(\phi_1^2\gamma_{dd}+\phi_2^2)\big\}\phi_1^2\dot{\gamma}_{dd}.
\end{align*}
The expectation of $y_{dt}^2$ is $\mathbb{E}_{\ttheta}[y_{dt}^2]=\mathbb{E}_v\big[\mathbb{E}_{\ttheta}[y_{dt}^2|\vv]\big]$, where
\begin{equation*}
\mathbb{E}_{\ttheta}[y_{dt}^2|\vv]=\mbox{var}_{\ttheta}[y_{dt}|\vv]+ \mathbb{E}_{\ttheta}^2[y_{dt}|\vv]=\nu_{dt}p_{dt}+\nu_{dt}^2p_{dt}^2.
\end{equation*}
Therefore
\begin{align*}
\mathbb{E}_{\ttheta}[y_{dt}^2]=\mathbb{E}_v\big[\mathbb{E}_{\ttheta}[y_{dt}^2|\vv]\big]&=
\int_{-\infty}^{\infty}\int_{-\infty}^{\infty}\nu_{dt}p_{dt}f(v_{1,d})f(v_{2,dt})\,dv_{1,d}dv_{2,dt}
\\[.15cm]
&+
\int_{-\infty}^{\infty}\int_{-\infty}^{\infty}\nu_{dt}^2p_{dt}^2f(v_{1,d})f(v_{2,dt})\,dv_{1,d}dv_{2,dt}
=
S_1+S_2,
\end{align*}
where
\begin{align*}
S_2&=\int_{-\infty}^{\infty}\int_{-\infty}^{\infty}\nu_{dt}^2p_{dt}^2f(v_{2,dt})f(v_{1,d})\,dv_{2,dt}dv_{1,d}
\\[.15cm]
&=
\nu_{dt}^2\int_{-\infty}^{\infty}\left[ \int_{-\infty}^{\infty}
\exp\big\{2\xx_{dt}\bbeta+2\phi_1 v_{1,d}+2\phi_2 v_{2,dt}\big\}f(v_{2,dt})\,dv_{2,dt}\right] f(v_{1,d})\,dv_{1,d}
\\[.15cm]
&=\nu_{dt}^2\int_{-\infty}^{\infty}
\exp\big\{2(\xx_{dt}\bbeta+\phi_2^2)+2\phi_1v_{1,d}\big\}f(v_{1,d})\,dv_{1,d}
\\[.15cm]
&=\nu_{dt}^2\exp\big\{2(\xx_{dt}\bbeta+\phi_1^2\gamma_{dd}+\phi_2^2)\big\}.
\end{align*}
Then, the expectation $\mathbb{E}_{\ttheta}[y_{dt}^2]$ is
\begin{equation*}
\mathbb{E}_{\ttheta}[y_{dt}^2]=\nu_{dt}\exp\big\{\xx_{dt}\bbeta+\frac12(\phi_1^2\gamma_{dd}+\phi_2^2)\big\}
+\nu_{dt}^2\exp\big\{2(\xx_{dt}\bbeta+\phi_1^2\gamma_{dd}+\phi_2^2)\big\},
\end{equation*}
and as a consequence, the $(p+2)$-th MM equation is
\begin{align*}
f_{p+2}(\ttheta)&=\frac{1}{DT}\sum_{d=1}^D\sum_{t=1}^T\bigg\{ \nu_{dt}\exp\Big\{ \xx_{dt}\bbeta+\frac12(\phi_1^2\gamma_{dd}+\phi_2^2)\Big\}
+ \nu_{dt}^2\exp\big\{ 2(\xx_{dt}\bbeta+\phi_1^2\gamma_{dd}+\phi_2^2)\big\}\bigg\}
\\[.15cm]
&- \frac{1}{DT}\sum_{d=1}^D\sum_{t=1}^T y_{dt}^2.
\end{align*}
The derivatives of $\mathbb{E}_{\ttheta}[y_{dt}^2]$ are
\begin{align*}
\frac{\partial \mathbb{E}_{\ttheta}[y_{dt}^2]}{\partial\beta_k}&=
\nu_{dt}\exp\big\{\xx_{dt}\bbeta+\frac12(\phi_1^2\gamma_{dd}+\phi_2^2)\big\}x_{dtk}
+
2\nu_{dt}^2\exp\big\{2(\xx_{dt}\bbeta+\phi_1^2\gamma_{dd}+\phi_2^2)\big\}x_{dtk},
\\[.15cm]
\frac{\partial \mathbb{E}_{\ttheta}[y_{d}^2]}{\partial\phi_1}&=
\nu_{dt}\exp\big\{\xx_{dt}\bbeta+\frac12(\phi_1^2\gamma_{dd}+\phi_2^2)\big\}\phi_1\gamma_{dd}
+
4\nu_{dt}^2\exp\big\{2(\xx_{dt}\bbeta+\phi_1^2\gamma_{dd}+\phi_2^2)\big\}\phi_1\gamma_{dd},
\\[.15cm]
\frac{\partial \mathbb{E}_{\ttheta}[y_{d}^2]}{\partial\phi_2}&=
\nu_{dt}\exp\big\{\xx_{dt}\bbeta+\frac12(\phi_1^2\gamma_{dd}+\phi_2^2)\big\}\phi_2
+
4\nu_{dt}^2\exp\big\{2(\xx_{dt}\bbeta+\phi_1^2\gamma_{dd}+\phi_2^2)\big\}\phi_2,
\\[.15cm]
\frac{\partial \mathbb{E}_{\ttheta}[y_{d}^2]}{\partial\rho}&=
\frac12\nu_{dt}\exp\big\{\xx_{dt}\bbeta+\frac12(\phi_1^2\gamma_{dd}+\phi_2^2)\big\}\phi_1^2\dot{\gamma}_{dd}
+
2\nu_{dt}^2\exp\big\{2(\xx_{dt}\bbeta+\phi_1^2\gamma_{dd}+\phi_2^2)\big\}\phi_1^2\dot{\gamma}_{dd}.
\end{align*}
The expectation of $y_{d.}^2$ is $\mathbb{E}_{\ttheta}[y_{d.}^2]=\mathbb{E}_v\big[\mathbb{E}_{\ttheta}[y_{d.}^2|\vv]\big]$, where
\begin{equation*}
y_{d.}^2=\sum_{t=1}^{T}y_{dt}^2+\sum_{t_1\neq t_2}y_{dt_1}y_{dt_2},\quad
\mathbb{E}_{\ttheta}[y_{dt}^2|\vv]=\mbox{var}_{\ttheta}[y_{dt}|\vv]+\mathbb{E}_{\ttheta}^2[y_{dt}|\vv]=\nu_{dt}p_{dt}+\nu_{dt}^2p_{dt}^2.
\end{equation*}
The expectation of $y_{d.}^2$, conditionally on the random effects $\vv$, is
\begin{align*}
\mathbb{E}_{\ttheta}[y_{d.}^2|\vv] &=
\sum_{t=1}^T \mathbb{E}_{\ttheta}[y_{dt}^2|\vv]+\sum_{t_1\neq t_2}\mathbb{E}_{\ttheta}[y_{dt_1}|\vv] \mathbb{E}_{\ttheta}[y_{dt_2}|\vv]
\\
&=
\sum_{t=1}^T\big\{\nu_{dt}p_{dt}+\nu_{dt}^2p_{dt}^2\big\}+\sum_{t_1\neq t_2}\nu_{dt_1}p_{dt_1}\nu_{dt_2}p_{dt_2}.
\end{align*}
Therefore
\begin{equation*}
\mathbb{E}_{\ttheta}[y_{d.}^2]=
\sum_{t=1}^T\nu_{dt} \mathbb{E}_v[p_{dt}]+\sum_{t=1}^T\nu_{dt}^2 \mathbb{E}_v[p_{dt}^2]
+\sum_{t_1\neq t_2}\nu_{dt_1}\nu_{dt_2} \mathbb{E}_v[p_{dt_1}p_{dt_2}],
\end{equation*}
where the expectation of $p_{dt_1}p_{dt_2}$ is
\begin{align*}
\mathbb{E}_v[p_{dt_1}p_{dt_2}]&=
\int_{-\infty}^{\infty}\int_{-\infty}^{\infty}\int_{-\infty}^{\infty}
\exp\{(\xx_{dt_1}+\xx_{dt_2})\bbeta+2\phi_1v_{1,d}+\phi_2v_{2,dt_1}+\phi_2v_{2,dt_2}\}
\\
&\cdot f(v_{2,dt_1})f(v_{2,dt_2})f(v_{1,d})\,dv_{2,dt_1}dv_{2,dt_2}dv_{1,d}
\\
&=
\int_{-\infty}^{\infty}\int_{-\infty}^{\infty}
\exp\{(\xx_{dt_1}+\xx_{dt_2})\bbeta+\frac12\phi_2^2+\phi_2v_{2,dt_1}+2\phi_1v_{1,d}\}
\\
&\cdot
f(v_{2,dt_1})f(v_{1,d})\,dv_{2,dt_1}dv_{1,d}
\\
&= \int_{-\infty}^{\infty}
\exp\{(\xx_{dt_1}+\xx_{dt_2})\bbeta+\frac12\phi_2^2+\frac12\phi_2^2+2\phi_1v_{1,d}\}f(v_{1,d})\,dv_{1,d}
\\
&= \exp\{(\xx_{dt_1}+\xx_{dt_2})\bbeta+\phi_2^2+2\phi_1^2\gamma_{dd}\}.
\end{align*}
Then, the expectation of $y_{d.}^2$ is
\begin{align*}
\mathbb{E}_{\ttheta}[y_{d.}^2]&=
\sum_{t=1}^T\nu_{dt}\exp\big\{\xx_{dt}\bbeta+\frac12(\phi_1^2\gamma_{dd}+\phi_2^2)\big\}
+
\sum_{t=1}^T\nu_{dt}^2\exp\big\{2\xx_{dt}\bbeta+2(\phi_1^2\gamma_{dd}+\phi_2^2)\big\}
\\
&+\sum_{t_1\neq t_2}\nu_{dt_1}\nu_{dt_2}\exp\{(\xx_{dt_1}+\xx_{dt_2})\bbeta+2\phi_1^2\gamma_{dd}+\phi_2^2\}
\\
&\pm
\sum_{t=1}^T\nu_{dt}^2\exp\big\{2\xx_{dt}\bbeta+2\phi_1^2\gamma_{dd}+\phi_2^2\big\}
\\
&=\sum_{t=1}^T\nu_{dt}\exp\big\{\xx_{dt}\bbeta+\frac12(\phi_1^2\gamma_{dd}+\phi_2^2)\big\}
+
\sum_{t=1}^T\nu_{dt}^2\exp\big\{2\xx_{dt}\bbeta+2(\phi_1^2\gamma_{dd}+\phi_2^2)\big\}
\\
&-\sum_{t=1}^T\nu_{dt}^2\exp\big\{2\xx_{dt}\bbeta+2\phi_1^2\gamma_{dd}+\phi_2^2\big\}
+
\left(\sum_{t=1}^T\nu_{dt}\exp\big\{\xx_{dt}\bbeta+\phi_1^2\gamma_{dd}+\frac{1}{2}\phi_2^2\big\}\right)^2,
\end{align*}
and as a consequence, the $(p+1)$-th MM equation is
\begin{align*}
f_{p+1}(\ttheta)&=\frac{1}{D}\sum_{d=1}^D \left\{ \vphantom{\left(\sum_{t=1}^T\nu_{dt}\exp\big\{\xx_{dt}\bbeta+\phi_1^2\gamma_{dd}+\frac{1}{2}\phi_2^2\big\}\right)^2}
\sum_{t=1}^T\nu_{dt}\exp\big\{\xx_{dt}\bbeta+\frac12(\phi_1^2\gamma_{dd}+\phi_2^2)\big\}
\right.
\\
&+ \left.
\big(e^{\phi_2^2}-1\big)\sum_{t=1}^T\nu_{dt}^2\exp\big\{2\xx_{dt}\bbeta+2\phi_1^2\gamma_{dd}+\phi_2^2)\big\}\right.
\\
&+\left.
\left(\sum_{t=1}^T\nu_{dt}\exp\big\{\xx_{dt}\bbeta+\phi_1^2\gamma_{dd}+\frac{1}{2}\phi_2^2\big\}\right)^2\right\}
-
\frac{1}{D}\sum_{d=1}^Dy_{d.}^2.
\end{align*}
The derivatives of $\mathbb{E}_{\ttheta}[y_{d.}^2]$ are
\begin{eqnarray*}
\frac{\partial \mathbb{E}_{\ttheta}[y_{d.}^2]}{\partial\beta_k}&=&
\sum_{t=1}^T\nu_{dt}P_{dt}x_{dtk}
+
2\sum_{t=1}^T\nu_{dt}^2Q_{dt} x_{dtk}
-2\sum_{t=1}^T\nu_{dt}^2R_{dt}x_{dtk}
\\
&+& 2\bigg(\sum_{t=1}^T\nu_{dt}S_{dt} \bigg)
\sum_{t=1}^T\nu_{dt}S_{dt}x_{dtk},
\\
\frac{\partial \mathbb{E}_{\ttheta}[y_{d.}^2]}{\partial\phi_1}&=&
\sum_{t=1}^T\nu_{dt}P_{dt}\phi_1\gamma_{dd}
+ 4\sum_{t=1}^T\nu_{dt}^2 Q_{dt}\phi_1\gamma_{dd}
- 4\sum_{t=1}^T\nu_{dt}^2 R_{dt}\phi_1\gamma_{dd}
\\
&+&4\bigg(\sum_{t=1}^T\nu_{dt}S_{dt}\bigg)
\sum_{t=1}^T\nu_{dt}S_{dt}\phi_1\gamma_{dd},
\\
\frac{\partial \mathbb{E}_{\ttheta}[y_{d.}^2]}{\partial\phi_2}&=&
\sum_{t=1}^T\nu_{dt}P_{dt}\phi_2
+ 4\sum_{t=1}^T\nu_{dt}^2 Q_{dt}\phi_2
- 2\sum_{t=1}^T\nu_{dt}^2 R_{dt}\phi_2
\\
&+& 2\bigg(\sum_{t=1}^T\nu_{dt}S_{dt}\bigg)
\sum_{t=1}^T\nu_{dt}S_{dt}\phi_2,
\\
\frac{\partial \mathbb{E}_{\ttheta}[y_{d.}^2]}{\partial\rho}&=&
\sum_{t=1}^T\frac12\nu_{dt}P_{dt}\phi_1^2\dot{\gamma}_{dd}
+ 2\sum_{t=1}^T\nu_{dt}^2 Q_{dt}\phi_1^2\dot{\gamma}_{dd}
- 2\sum_{t=1}^T\nu_{dt}^2 R_{dt}\phi_1^2\dot{\gamma}_{dd}
\\
& +&
2\bigg(\sum_{t=1}^T\nu_{dt} S_{dt} \bigg)
\sum_{t=1}^T \nu_{dt} S_{dt}\phi_1^2\dot{\gamma}_{dd},
\end{eqnarray*}
where
\begin{eqnarray*}
P_{dt} &=& \exp\big\{\xx_{dt}\bbeta+\frac12(\phi_1^2\gamma_{dd}+\phi_2^2)\big\},\quad
Q_{dt} = \exp\big\{2\xx_{dt}\bbeta+2(\phi_1^2\gamma_{dd}+\phi_2^2)\big\},
\\
R_{dt} &=& \exp\big\{2\xx_{dt}\bbeta+2\phi_1^2\gamma_{dd}+\phi_2^2\big\},\quad
S_{dt} = \exp\big\{\xx_{dt}\bbeta+\phi_1^2\gamma_{dd}+\frac{1}{2}\phi_2^2\big\}.
\end{eqnarray*}
The expectation of $y_{d_1.}y_{d_2.}$ is
\begin{align*}
\mathbb{E}_{\ttheta}[y_{d_1.}y_{d_2.}]&= \mathbb{E}_v\big[\mathbb{E}_{\ttheta}[y_{d_1.}y_{d_2.}|\vv]\big]=
\mathbb{E}_v\big[\mathbb{E}_{\ttheta}[y_{d_1.}|v_{1,d_1},\vv_{2,d_1}]\mathbb{E}_{\ttheta}[y_{d_2.}|v_{1,d_2},\vv_{2,d_2}]\big]
\\[.15cm]
&=
\sum_{t_1=1}^T\sum_{t_2=1}^T \mathbb{E}_v\big[\mathbb{E}_{\ttheta}[y_{d_1t_1}|v_{1,d_1},v_{2,d_1t_1}]\mathbb{E}_{\ttheta}[y_{d_2t_2}|v_{1,d_2},v_{2,d_2t_2}]\big]
\\[.15cm]
&=
\sum_{t_1=1}^T\sum_{t_2=1}^T\nu_{d_1t_1}\nu_{d_2t_2}\mathbb{E}_v\big[p_{d_1t_1}p_{d_2t_2}\big].
\end{align*}
By defining $\varphi_{d_1d_2}^{t_1t_2}(\ttheta)= \mathbb{E}_v\big[p_{d_1t_1}p_{d_2t_2}\big]$, it holds that
\begin{align*}
\varphi_{d_1d_2}^{t_1t_2}(\ttheta)&=
\int_{\R^4}
\exp\big\{(\xx_{d_1t_1}+\xx_{d_2t_2})\bbeta+\phi_1(v_{1,d_1}+v_{1,d_2})
+ \phi_2(v_{2,d_1t_1}+v_{2,d_2t_2})\big\}
\\[.15cm]
&\cdot
f(v_{2,d_2t_2})dv_{2,d_2t_2}\,f(v_{2,d_1t_1})dv_{2,d_1t_1}
f(v_{1,d_2}|v_{1,d_1})dv_{1,d_2}\,f(v_{1,d_1})\,dv_{1,d_1}
\\[.15cm]
&=
\int_{\R^3}
\exp\big\{(\xx_{d_1t_1}+\xx_{d_2t_2})\bbeta+\phi_1(v_{1,d_1}+v_{1,d_2})+\phi_2v_{2,d_1t_1}+\frac{1}{2}\phi_2^2\big\}
\\[.15cm]
&\cdot
f(v_{2,d_1t_1})dv_{2,d_1t_1}\,f(v_{1,d_2}|v_{1,d_1})dv_{1,d_2}\,f(v_{1,d_1})\,dv_{1,d_1}.
\end{align*}
Therefore,
\begin{align*}
\varphi_{d_1d_2}^{t_1t_2}(\ttheta)&=
\int_{\R^2}
\exp\big\{(\xx_{d_1t_1}+\xx_{d_2t_2})\bbeta+\phi_1(v_{1,d_1}+v_{1,d_2})+\frac{1}{2}\phi_2^2+\frac{1}{2}\phi_2^2\big\}
\\[.15cm]
&\cdot
f(v_{1,d_2}|v_{1,d_1})dv_{1,d_2}\,f(v_{1,d_1})\,dv_{1,d_1}
\\[.15cm]
&=
\int_{\R}
\exp\Big\{(\xx_{d_1t_1}+\xx_{d_2t_2})\bbeta+\phi_1v_{1,d_1}+\frac{\gamma_{d_1d_2}}{\gamma_{d_1d_1}}v_{1,d_1}\phi_1
\\[.15cm]
& +
\frac12\Big(\gamma_{d_2d_2}-\frac{\gamma_{d_1d_2}^2}{\gamma_{d_1d_1}}\Big)\phi_1^2+\phi_2^2\Big\}
f(v_{1,d_1})\,dv_{1,d_1}
\\[.15cm]
&=
\exp\Big\{(\xx_{d_1t_1}+\xx_{d_2t_2})\bbeta+\frac12\Big(1+\frac{\gamma_{d_1d_2}}{\gamma_{d_1d_1}}\Big)^2\gamma_{d_1d_1}\phi_1^2
+
\frac12\Big(\gamma_{d_2d_2}-\frac{\gamma_{d_1d_2}^2}{\gamma_{d_1d_1}}\Big)\phi_1^2+\phi_2^2\Big\}
\\[.15cm]
&=\exp\Big\{(\xx_{d_1t_1}+\xx_{d_2t_2})\bbeta+\frac12\,\phi_1^2(\gamma_{d_1d_1}+2\gamma_{d_1d_2}+\gamma_{d_2d_2})+\phi_2^2\Big\}.
\end{align*}
Therefore, the $(p+3)$-th MM equation is
\begin{align*}
f_{p+3}(\ttheta)&=\frac{1}{D(D-1)}\sum_{d_1\neq d_2}^D\sum_{t_1=1}^T\sum_{t_2=1}^T\nu_{d_1t_1}\nu_{d_2t_2}\exp\Big\{(\xx_{d_1t_1}+\xx_{d_2t_2})\bbeta
\\
&+\frac12\,\phi_1^2(\gamma_{d_1d_1}+2\gamma_{d_1d_2}+\gamma_{d_2d_2})+\phi_2^2\Big\} - \frac{1}{D(D-1)}\sum_{d_1\neq d_2}^D\sum_{t_1=1}^T\sum_{t_2=1}^T\nu_{d_1t_1}y_{d_1.}y_{d_2.}
\end{align*}
The derivatives of $\varphi_{d_1d_2}^{t_1t_2}(\ttheta)$ are
\begin{align*}
\frac{\partial \varphi_{d_1d_2}^{t_1t_2}(\ttheta)}{\partial\beta_k}&=
\varphi_{d_1d_2}^{t_1t_2}(\ttheta)(x_{d_1t_1k}+x_{d_2t_2k}),
\\
\frac{\partial \varphi_{d_1d_2}^{t_1t_2}(\ttheta)}{\partial\phi_1}&=
\varphi_{d_1d_2}^{t_1t_2}(\ttheta)\phi_1(\gamma_{d_1d_1}+\gamma_{d_2d_2}+2\gamma_{d_1d_2}),
\\
\frac{\partial \varphi_{d_1d_2}^{t_1t_2}(\ttheta)}{\partial\phi_2}&=
2\varphi_{d_1d_2}^{t_1t_2}(\ttheta)\phi_2,
\\
\frac{\partial \varphi_{d_1d_2}^{t_1t_2}(\ttheta)}{\partial\rho}&=
\frac12\varphi_{d_1,d_2}(\ttheta)\phi_1^2(\dot{\gamma}_{d_1d_1}+\dot{\gamma}_{d_2d_2}+2\dot{\gamma}_{d_1d_2}).
\end{align*}
The elements of the Jacobian matrix are
\begin{align*}
H_{kr}&=\frac{\partial f_k(\ttheta)}{\partial \theta_r}=
\frac{1}{DT}\sum_{d=1}^D\sum_{t=1}^{T}\frac{\partial \mathbb{E}_{\ttheta}[y_{dt}]}{\partial\theta_r}\,x_{dtk},\quad k=1,\ldots,p,\,
r=1,\ldots,p+3,
\\
H_{p+1r}&=\frac{\partial f_{p+1}(\ttheta)}{\partial \theta_r}=
\frac{1}{D}\sum_{d=1}^D\frac{\partial \mathbb{E}_{\ttheta}[y_{d.}^2]}{\partial\theta_r},\quad r=1,\ldots,p+3,
\\
H_{p+2r}&=\frac{\partial f_{p+2}(\ttheta)}{\partial \theta_r}=
\frac{1}{DT}\sum_{d=1}^D\sum_{t=1}^T\frac{\partial \mathbb{E}_{\ttheta}[y_{dt}^2]}{\partial\theta_r},\quad r=1,\ldots,p+3,
\\
H_{p+3r}&=\frac{\partial f_{p+3}(\ttheta)}{\partial \theta_r}=
\frac{1}{D(D-1)}\sum_{d_1\neq d_2}^D\sum_{t_1=1}^T\sum_{t_2=1}^T\nu_{d_1t_1}\nu_{d_2t_2}\frac{\partial \varphi_{d_1d_2}^{t_1t_2}(\ttheta)}{\partial\theta_r},\quad
r=1,\ldots,p+3.
\end{align*}
\section{Appendix: Simulations}\label{apsim}

This section presents two model-based simulation experiments.
The first one studies the the behaviour of the MM fitting algorithm while the second one compares the performance of the two introduced predictors, i.e.
the EBP and the plug-in. The response variables are generated independently as $y_{dt}\vert v_{1, d}, v_{2, dt} \sim \mbox{Poisson}(\nu_{dt}p_{dt})$, where
\begin{equation*}
p_{dt}=\exp\{\beta_0+x_{dt}\beta_1+\phi_1 v_{1,d}+\phi_2 v_{2,dt}\}, \,
x_{dt}=\frac{d+t/T}{D},\, d=1,\ldots, D, \, t=1,\ldots, T.
\end{equation*}
The domain random effects, $v_{1, d}\, (d=1,\ldots, D)$, are generated according to a SAR(1) process, i.e.
\begin{equation*}
\vv_1 = \underset{1\leq d \leq D}{\mbox{col}}(v_{1,d})=(\II_D-\rho \WW)^{-1}\uu_1,
\end{equation*}
where $\II_D$ denotes the $D\times D$ identity matrix, $\rho$ is the autocorrelation parameter, $\WW=(\omega_{ij})_{i,j=1,\ldots, D}$ is a proximity matrix and $\uu_1 \sim N(\cero, \II_D)$. For the $D\times D$ proximity matrix $\WW$, a $7$-diagonal matrix is considered.
Let $k$ be the number of diagonals of $\WW$, then the number of upper and lower diagonals is $m=\lfloor k/2\rfloor$, where $\lfloor k/2\rfloor$ denotes integer part of $k/2$. The diagonals are denoted by $1$ (main) and $j$ (upper and lower), $j=2,\ldots,m+1$. The diagonals are constructed in the following way.
\begin{itemize}
\item
Diagonal 1 (main): if $|i-j|=0$, then $\omega_{ij}=0$.
\item
Diagonal 2 (upper and lower):  if $|i-j|=1$, then $\omega_{ij}=\frac{1}{2}-\big(\frac{1}{2^3} + \frac{1}{2^4}\big)$.
\item
Diagonals $3-(m+1)$ (upper and lower): if $|i-j|\in\{2,\ldots,m\}$, then $\omega_{ij}=1/2^{|i-j|+1}$.
\end{itemize}
Then, the elements of diagonals 1, 2, 3 and 4 (upper and lower) are $0$, $5/16$, $2/16$ and $1/16$ respectively. This rule does not apply to the first and the last $m$ rows ($m=3$). For those rows, the numerators of the diagonal elements are kept fixed ($0$, $5$, $2$ and $1$) and the denominators are recalculated so that the sum of the row is $1$. Using this criterion, the $9\times 9$ $7$-diagonal matrix $\WW$ is
\begin{equation*}\label{WW7}
\WW=\left(\begin{array}{ccccccccc}
0& 5/8& 2/8& 1/8& 0& 0&0&0&0\\
5/13& 0& 5/13& 2/13& 1/13& 0&0&0&0\\
2/15& 5/15& 0& 5/15& 2/15& 1/15&0&0&0\\
\hline
1/16& 2/16& 5/16& 0& 5/16& 2/16&1/16&0&0\\
0& 1/16& 2/16& 5/16& 0& 5/16& 2/16&1/16&0\\
0& 0&1/16& 2/16& 5/16& 0& 5/16& 2/16&1/16\\
\hline
0&0& 0& 1/15& 2/15& 5/15& 0&5/15& 2/15\\
0&0& 0& 0& 1/13& 2/13& 5/13& 0&5/13\\
0&0& 0& 0& 0& 1/8& 2/8& 5/8& 0\\
\end{array}\right).
\end{equation*}
That $9\times 9$ $7$-diagonal matrix $\WW$ can be generalized to a $D\times D$ matrix by repeating $D-6$ times the weights of the central rows (i.e., $0$, $5/16$, $2/16$ and $1/16$).

In both simulation experiments, we take $\beta_0=-3$, $\beta_1=0.8$, $\phi_1=0.5$, $\phi_2=0.5$ and $\nu_{dt}=100$, $d=1,\ldots,D$, $t=1,\ldots,T$. The simulation considers the scenarios
$D=100$ and $T=4,8$ for studying the influence of the time periods.
For each scenario, it takes $\rho=0.1, 0.3, 0.5$.
This section runs the Monte Carlo simulation experiments with $K=1000$ iterations.

%
\subsection{Simulation 1}\label{sim1}
%
The target of Simulation 1 is to check the behaviour of the MM fitting algorithm introduced in Section \ref{sec2}.
Table \ref{sec5-tabSim1.1} presents the bias and Table \ref{sec5-tabSim1.2} the root mean squared error (RMSE) for the model parameters $\theta \in \ttheta = \{ \beta_0, \beta_1, \phi_1, \phi_2, \rho \}$. This section considers two options for estimating the vector of all model parameters $\ttheta$. In the first option (Opt. 1), the vector $\hat{\ttheta}$ is obtained as a solution of the system of $p+3$ nonlinear equations (\ref{secMM-MM}). In the second option (Opt. 2), $\hat{\rho}$ is given by calculating the Moran's I measure over the predicted domain random effects under Model T1.
\renewcommand{\arraystretch}{1.2}
\begin{table}[H]
\caption{Bias of the MM fitting algorithm under Model ST1.}
\centering
\begin{tabular}{llrrrrr}
\toprule
& & \multicolumn{2}{c}{$T=4$} & & \multicolumn{2}{c}{$T=8$}\\
\cmidrule{3-4} \cmidrule{6-7}
$\rho$ & &  \multicolumn{1}{c}{Opt. 1} & \multicolumn{1}{c}{Opt. 2} & & \multicolumn{1}{c}{Opt. 1} & \multicolumn{1}{c}{Opt. 2}\\
\midrule
0.1 & $\hat{\beta}_0$ & 0.0115 & 0.0115 && 0.0171 & 0.0171 \\
    & $\hat{\beta}_1$ &-0.0145 &-0.0145 &&-0.0233 &-0.0234 \\
    & $\hat{\phi}_1$  &-0.0219 &-0.0230 &&-0.0240 &-0.0250 \\
    & $\hat{\phi}_2$  &-0.0098 &-0.0098 &&-0.0066 &-0.0066 \\
    & $\hat{\rho}$    &-0.1689 &-0.0848 &&-0.1593 &-0.0820 \\
\midrule
0.3 & $\hat{\beta}_0$ & 0.0196 & 0.0197 && 0.0150 & 0.0151 \\
    & $\hat{\beta}_1$ &-0.0225 &-0.0226 &&-0.0168 &-0.0169 \\
    & $\hat{\phi}_1$  &-0.0144 &-0.0186 &&-0.0129 &-0.0181 \\
    & $\hat{\phi}_2$  &-0.0099 &-0.0099 &&-0.0087 &-0.0087 \\
    & $\hat{\rho}$    &-0.3236 &-0.1986 &&-0.3441 &-0.1865 \\
\midrule
0.5 & $\hat{\beta}_0$ & 0.0285 & 0.0285 && 0.0394 & 0.0394 \\
    & $\hat{\beta}_1$ &-0.0387 &-0.0387 &&-0.0666 &-0.0664 \\
    & $\hat{\phi}_1$  & 0.0101 & 0.0059 && 0.0062 & 0.0033 \\
    & $\hat{\phi}_2$  &-0.0123 &-0.0123 &&-0.0111 &-0.0111 \\
    & $\hat{\rho}$    &-0.7090 &-0.2856 &&-0.8400 &-0.2601 \\
\bottomrule
\end{tabular}
\label{sec5-tabSim1.1}
\end{table}

Both options behave similarly for the fixed effects and the variance parameters. For these parameters, the variance is the most important term of the MSE since bias is much smaller than RMSE. On the other hand, Opt. 2 produces more competitive estimates for the autocorrelation parameter $\rho$, since it drastically reduces both bias and RMSE. For $\rho$, bias is the main part of the MSE since it takes similar absolute values to the RMSE. Then, a bias correction by bootstrap might be useful.

\renewcommand{\arraystretch}{1.2}
\begin{table}[H]
\caption{Root mean squared error of the MM fitting algorithm under Model ST1.}
\centering
\begin{tabular}{llrrrrr}
\toprule
& & \multicolumn{2}{c}{$T=4$} & & \multicolumn{2}{c}{$T=8$}\\
\cmidrule{3-4} \cmidrule{6-7}
$\rho$ & &  \multicolumn{1}{c}{Opt. 1} & \multicolumn{1}{c}{Opt. 2} & & \multicolumn{1}{c}{Opt. 1} & \multicolumn{1}{c}{Opt. 2}\\
\midrule
0.1 & $\hat{\beta}_0$ & 0.1388 & 0.1387 && 0.1291 & 0.1291 \\
    & $\hat{\beta}_1$ & 0.2337 & 0.2337 && 0.2162 & 0.2161 \\
    & $\hat{\phi}_1$  & 0.0658 & 0.0659 && 0.0561 & 0.0563 \\
    & $\hat{\phi}_2$  & 0.0449 & 0.0449 && 0.0289 & 0.0289 \\
    & $\hat{\rho}$    & 0.1904 & 0.1129 && 0.1796 & 0.1090 \\
\midrule
0.3 & $\hat{\beta}_0$ & 0.1682 & 0.1682 && 0.1554 & 0.1554 \\
    & $\hat{\beta}_1$ & 0.2824 & 0.2823 && 0.2609 & 0.2609 \\
    & $\hat{\phi}_1$  & 0.0649 & 0.0647 && 0.0569 & 0.0545 \\
    & $\hat{\phi}_2$  & 0.0463 & 0.0463 && 0.0309 & 0.0309 \\
    & $\hat{\rho}$    & 0.3537 & 0.2144 && 0.3872 & 0.2039 \\
\midrule
0.5 & $\hat{\beta}_0$ & 0.2189 & 0.2189 && 0.2100 & 0.2100 \\
    & $\hat{\beta}_1$ & 0.3736 & 0.3736 && 0.3509 & 0.3509 \\
    & $\hat{\phi}_1$  & 0.0742 & 0.0650 && 0.0696 & 0.0530 \\
    & $\hat{\phi}_2$  & 0.0451 & 0.0451 && 0.0307 & 0.0307 \\
    & $\hat{\rho}$    & 0.7766 & 0.2998 && 0.9038 & 0.2755 \\
\bottomrule
\end{tabular}
\label{sec5-tabSim1.2}
\end{table}

%
\subsection{Simulation 2}\label{sim2}
%
The second simulation experiment investigates the behaviour of the considered $p_{dt}$ predictors for different time instants, $T$, and autocorrelation parameters, $\rho$. Specifically, it calculates BP-plug-in, BP, plug-in and EBP. Given the computational burden presented by the BPs (and EBPs) of the target parameter and of the two random effects under Model ST1, the simulation considers their approximated versions (see Section \ref{sec3}). The first predictor, BP-plug-in, is obtained from (\ref{plugMST1}) by using the theoretical vector of model parameters $\ttheta$.
The BPs and EBPs are approximated by generating $S1=500$ random variables $\vv_{1}^{(s_1)}$ and $S2=700$ random variables $v_{2,\ell\tau}^{(s2)}$. For the empirical predictors (plug-in and EBP), the model parameters are estimated by using the second option in MM, since it has presented better results in the previous simulation experiment.

Table \ref{sec5-tabSim2} presents the average across domains and time instants of the biases and the RMSEs (both $\times 10^2$) for BP-plug-in, BP, plug-in and EBP. BP and EBP are more competitive than the respective plug-in. Specially in bias, where a significant improvement is achieved. The obtained results also suggest that $T$ does not affect too much the results and that variance is the most important term of the MSE since bias is much smaller than RMSE.
\renewcommand{\arraystretch}{1}
\begin{table}[H]
\caption{Bias (B) and root mean squared error (RMSE) of the BP-plug-in, BP, plug-in and EBP of $p_{dt}$ under Model ST1 (both $\times 10^2$).}
\centering
\begin{tabular}{llrrrrr}
\toprule
& & \multicolumn{2}{c}{$T=4$} & & \multicolumn{2}{c}{$T=8$}\\
\cmidrule{3-4} \cmidrule{6-7}
$\rho$ & Predictors &  \multicolumn{1}{c}{B} & \multicolumn{1}{c}{RMSE} & & \multicolumn{1}{c}{B} & \multicolumn{1}{c}{RMSE}\\
\midrule
0.1 & BP-plug-in & 0.3350 & 2.8325 && 0.3318 & 2.7739 \\
    & BP         & 0.0701 & 2.7581 && 0.0689 & 2.7165 \\
    & plug-in    & 0.3251 & 2.8248 && 0.3261 & 2.7710 \\
    & EBP        & 0.0699 & 2.7645 && 0.0687 & 2.7203 \\
\midrule
0.3 & BP-plug-in & 0.3423 & 2.8445 && 0.3217 & 2.7840 \\
    & BP         & 0.0711 & 2.7655 && 0.0674 & 2.7306 \\
    & plug-in    & 0.3406 & 2.8354 && 0.3194 & 2.7851 \\
    & EBP        & 0.0717 & 2.7723 && 0.0677 & 2.7342 \\
\midrule
0.5 & BP-plug-in & 0.3363 & 2.8939 && 0.3311 & 2.8125 \\
    & BP         & 0.0692 & 2.8129 && 0.0829 & 2.7562 \\
    & plug-in    & 0.3290 & 2.8830 && 0.3282 & 2.8169 \\
    & EBP        & 0.0699 & 2.8200 && 0.0824 & 2.7595 \\
\bottomrule
\end{tabular}
\label{sec5-tabSim2}
\end{table}

The system of MM nonlinear equations (\ref{secMM-MM}) is solved by using the \textsf{R} \textit{nleqslv} package.
The \textit{mvtnorm} package is also used to generate samples of a SAR(1) process. The computational burden of the first option in MM is much higher.
Taking $T=4$, the average runtime of the first option is $60.4$ seconds, while for the second option is $0.1$ seconds.
On the other hand, regarding the computational burden of the $p_{dt}$ predictors, the EBP is faster than the plug-in.
The reason is because the proposed plug-in predictor requires the calculation of two EBPs ($\hat{\vv}_1$ and $\hat{\vv}_2$).
The average runtimes are $210.7$ seconds for the EBP and $320.5$ seconds for the plug-in.
\section*{References}
\begin{description}\setlength\itemsep{0em}
\item
Baldermann, C., Salvati, N., and Schmid, T. (2016). Robust small area estimation under
spatial non-stationarity. Discussion Paper, School of Business and Economics: Economics,
22, N. 2016/5.
\item
Boubeta, M., Lombard\'{\i}a, M. J., W., Marey-P\'erez, M., Morales, D. (2015).
Prediction of forest fires occurrences with area-level Poisson mixed models.
\textit{Journal of Environmental Management},  154, 151-158.
\item
Boubeta, M., Lombard\'{\i}a, M. J., and Morales, D. (2016). Empirical best prediction under
area-level Poisson mixed models. \textit{Test}, 25, 548-569.
\item
Boubeta, M., Lombard\'{\i}a, M. J., and Morales, D.(2017). Poisson mixed models for studying the poverty in small areas.
\textit{Computational Statistics and Data Analysis}, 107, 32-47.
\item
Boubeta, M., Lombard\'{\i}a, M. J., W., Marey-P\'erez, M., Morales, D. (2019). Poisson mixed models for predicting number of fires.
{\it International Journal of Wildland Fire}, 28, 3, 237-253.
\item
Chandra, H., Salvati, N., Chambers, R., and Tzavidis, N. (2012). Small area estimation under
spatial nonstationarity. \textit{Computational Statistics and Data Analysis}, 56, 2875-2888.
\item
Chandra, H., Salvati, N., and Chambers, R. (2015). A spatially nonstationary Fay-Herriot model for small area estimation. \textit{Journal of Survey Statistics and Methodology}, 3, 109-135.
\item
Chandra, H., Salvati, N., and Chambers, R. (2017). Small area prediction of counts under a non-stationary spatial model.
\textit{Spatial Statistics}, 20,  30-56.
\item
Chandra, H., Salvati, N., and Chambers, R. (2018). Small area estimation under a spatially non-linear model.
\textit{Computational Statistics and Data Analysis}, 126,  19-38.
\item
Choi, J., Lawson, A.B., Cai, B., Hossain, M.M. (2011). Evaluation of Bayesian spatiotemporal latent models in small area health data. {\it Environmetrics}, 22, 8, 1008-1022.
\item
Cressie, N. (1993). Statistics for spatial data. Wiley, New York.
\item
Esteban, M. D., Morales, D., P\'erez, A., L. Santamar\'{\i}a (2012). Small area estimation of poverty proportions under area-level time models.
{\it Computational Statistics and Data Analysis}, 56, 10, 2840-2855.
\item
Esteban, M. D., Morales, D., P\'erez, A. (2016). Area-level spatio-temporal small area
estimation models. In Analysis of poverty data by small area estimation (ed M. Pratesi).
John Wiley and Sons, Ltd, Chichester, UK.
\item
Esteban, M. D., Lombard\'{\i}a, M.J., L\'opez-Vizca\'{\i}no, E., Morales, D., P\'erez A. (2020).
Small area estimation of proportions under area-level compositional mixed models.
{\it TEST}.  DOI: 10.1007/s11749-019-00688-w.
\item
Gonz\'alez-Manteiga, W., Lombard\'{\i}a, M.J., Molina, I., Morales, D. and Santamar\'{\i}a, L. (2008).
Analytic and bootstrap approximations of prediction errors under a multivariate Fay-Herriot model.
\textit{Computational Statistics and Data Analysis}, 52, 5242-5252.
\item
Gonz\'alez-Manteiga, W., Lombard\'{\i}a, M. J., Molina, I., Morales, D., Santamar\'{\i}a, L.  (2010).
Small area estimation under Fay-Herriot models with nonparametric estimation of heteroscedasticity.
{\it Statistical Modelling}, 10,  215-239.
\item
Jiang, J. (1998). Consistent estimators in generalized linear models.
{\it Journal of the American Statistical Association}, {\bf  93}, 720-729.
\item
L\'opez-Vizca\'{\i}no, E., Lombard\'{\i}a, M.J., Morales, D. (2013).
Multinomial-based small area estimation of labour force indicators. {\it Statistical Modelling}, 13,  153-178.
\item[]
L\'opez-Vizca\'{\i}no, E., Lombard\'{\i}a, M.J. and Morales, D. (2015).
Small area estimation of labour force indicators under a multinomial model with correlated time and area effects.
\textit{Journal of the Royal Statistical Association, series A}, 178,  535-565.
\item
Marhuenda, Y., Molina, I., and Morales, D. (2013). Small area estimation with spatio-temporal
Fay-Herriot models. \textit{Computational Statistics and Data Analysis}, 58, 308 - 325. The Third
Special Issue on Statistical Signal Extraction and Filtering.
\item
Meteogalicia. Xunta de Galicia. In web site: \\ https://www.meteogalicia.gal/observacion/informesclima/informesIndex.action.
\item
Molina, I., Salvati, N., and Pratesi, M. (2009). Bootstrap for estimating the MSE of the spatial
EBLUP. \textit{Computational Statistics}, 24, 441-458.
\item
Moura, F. A. S. and Migon, H. S. (2002). Bayesian spatial models for small area estimation of
proportions. \textit{Statistical Modelling}, 2(3), 183-201.
\item
Opsomer, J. D., Claeskens, G., Ranalli, M. G., Kauermann, G., and Breidt, F. J. (2008).
Nonparametric small area estimation using penalized spline regression.
\textit{Journal of the Royal Statistical Society,  Series B}, 70, 265-286.
\item
Pereira, L.N., Coelho, P.S. (2012). Small area estimation using a spatio-temporal linear mixed model.
{\it REVSTAT - Statistical Journal}, 10, 285-308.
\item
Petrucci, A. and Salvati, N. (2006). Small area estimation for spatial correlation in watershed erosion assessment.
\textit{Journal of Agricultural, Biological, and Environmental Statistics}, 11, 169-172.
\item
Pratesi, M. and Salvati, N. (2008). Small area estimation: the EBLUP estimator based on
spatially correlated random area effects. \textit{Statistical Methods and Applications}, 17, 113-171.
\item
S\"arndal, C., Swensson, B., Wretman. J. (1992). Model assisted survey sampling. Springer
\item
Singh, B., Shukla, G., and Kundu, D. (2005). Spatio-temporal models in small area estimation.
\textit{Survey Methodology}, 31, 183-195.
\item
Sugasawa, S., Kawakubo, Y., and Ogasawara, K. (2015). Geographically weighted empirical
Bayes estimation via natural exponential family. Discussion Paper No. 2015-01. Tokyo Institute of Technology.
\item
Ugarte, M. D., Ib\'a\~nez, B., and Militino, A. F. (2006). Modelling risks in disease mapping.
\textit{Statistical methods in medical research}, 15, 21-35.
\item
Ugarte, M. D., Goicoa, T., and Militino, A. F. (2010). Spatio-temporal modeling of mortality
risk using penalized splines. \textit{Environmetrics}, 21, 270-289.
\item
You, Y. and Zhou, Q. M. (2011). Hierarchical Bayes small area estimation under a spatial
model with application to health survey data. \textit{Survey Methodology}, 37, 25-37.
\end{description}

\end{document}